\documentclass[times]{my-iapress}

\usepackage{moreverb}

\usepackage[colorlinks,bookmarksopen,bookmarksnumbered,citecolor=red,urlcolor=red]{hyperref}

\def\volumeyear{2019}
\def\volumenumber{7}
\def\volumemonth{June}
\usepackage{amsmath,amssymb}
\usepackage{graphicx}

\begin{document}

\runningheads{The risk of contagion spreading and its optimal control in the economy}{
O. Kostylenko, H. S. Rodrigues, D. F. M. Torres}

\title{The risk of contagion spreading and its optimal control in the economy\footnote[2]{This 
work is part of first author's Ph.D., which is carried out at University of Aveiro 
under the Doctoral Program in Applied Mathematics MAP-PDMA, 
of Universities of Minho, Aveiro and Porto.}}

\author{Olena Kostylenko\affil{1}$^,$\corrauth,
Helena Sofia Rodrigues\affil{1,2}, 
Delfim F. M. Torres\affil{1}}

\address{\affilnum{1}Center for Research and Development in Mathematics and Applications (CIDMA),
Department of Mathematics, University of Aveiro, 3810-193 Aveiro, Portugal.
\affilnum{2}School of Business Studies, Polytechnic Institute of Viana do Castelo,
4930-678 Valen\c{c}a, Portugal.}

\corraddr{Olena Kostylenko (Email: o.kostylenko@ua.pt). 
Center for Research and Development in Mathematics and Applications (CIDMA), 
Department of Mathematics, University of Aveiro, 3810-193 Aveiro, Portugal.}


\begin{abstract}
The global crisis of 2008 provoked a heightened interest among scientists 
to study the phenomenon, its propagation and negative consequences.
The process of modelling the spread of a virus is commonly used in epidemiology.
Conceptually, the spread of a disease among a population is similar to the contagion 
process in economy. This similarity allows considering the contagion 
in the world financial system using the same mathematical model of infection 
spread that is often used in epidemiology. Our research focuses on the dynamic 
behaviour of contagion spreading in the global financial network.
The effect of infection by a systemic spread of risks in the network 
of national banking systems of countries is tested.
An optimal control problem is then formulated to simulate a control 
that may avoid significant financial losses. The results show 
that the proposed approach describes well the reality 
of the world economy, and emphasizes the importance of international 
relations between countries on the financial stability.
\end{abstract}


\maketitle



\section{Introduction}

Formation of the world market without national barriers contributed to the beginning 
and further development of a fundamentally new global power system, worldwide.
This process, that is called globalization, is based on the growing interconnectedness 
and interdependence of the modern world, which characterizes the development 
of international relations in all spheres of the world community.
This leads to an increase in the role of external factors in the evolution
of all the countries participating in the process.

Scientists in the field of economics define the concept of globalization 
as an increase in the scale of capital movements, currency flows, foreign 
trade in goods and services, migration of people, and exchange of information, 
ideas and technologies. The World Bank defines globalization as the ``Freedom 
and ability of individuals and firms to initiate voluntary economic transactions 
with residents of other countries''. The International Monetary Fund (IMF) 
defines globalization as ``The growing economic interdependence of countries 
worldwide through increasing volume and variety of cross-border transactions 
in goods and services, freer international capital flows, and more rapid 
and widespread diffusion of technology'' \cite{IMF};
``The process by which the world becomes a single place'' \cite{King}.
Economic globalization is the increasing economic interdependence of national 
economies across the world through a rapid increase in cross-border movement 
of goods, services, technology, and capital \cite{Joshi}.
In this way, world globalization is a process of uniting the world into 
a single system of global property, where the global world economy is 
characterized by a close interconnection of economic development of different countries.

Synchronization of economic relations is constantly felt. However, as economic agents 
react more strongly to negative shocks than to positive ones, synchronization becomes 
more noticeable during a crisis. This is also facilitated by mass media, as they 
instantly make information about shocks available to a wide range of users.

The global crisis of 2008 has become a significant source of knowledge about the process 
of synchronization of economic relations during the crisis. It reveals 
that economically weakly connected countries in stable periods, demonstrate 
a unidirectional movement of macro indicators during the crisis period. 
This happens because the crisis in one country provokes a crisis in another(s).
This effect is termed \emph{financial contagion} and it is capable of generating a global 
network epidemic. Therefore, nowadays the prevention of the global economic crisis and risk 
minimization are important tasks in the field of economics.

The definition of contagion, which scientists consider in their works, 
varies considerably. However, there is a similarity in all considered models:
they always attempt to provide a structure, explaining why a shock in one country 
may be transmitted elsewhere \cite{Kaminsky}. Mechanisms of mutual influence 
of countries in crisis situations is studied using mathematical and 
instrumental methods of economic analysis. The main stage of modelling 
the epidemic of the financial crisis is the choice of a mathematical model that 
adequately describes the process.

Several researchers draw an analogy between economic and natural viruses, 
between network and biological epidemics. From a mathematical point 
of view, the process of spreading a financial virus in the economy 
is similar to the spread of a viral disease. Therefore, these processes 
are described by similar models of differential equations. Because economic 
interactions of countries form a network, the use of network 
theory can enrich our understanding of financial systems as complex systems, 
which is used to explain the process of virus contamination in the network.

Here, we simulate an epidemiological process of financial contagion 
using the Kermack and McKendrick SIR model \cite{Kermack}. The process 
of infection spreading among the agents is studied through network interconnections. 
Then, similarly to \cite{Kostylenko}, we apply optimal control theory
to minimize the negative consequences of the spread of the virus.

The remainder of this paper is divided into five sections. 
Section~\ref{sec:02} provides a detailed description of the data employed. 
Section~\ref{sec:03} presents the proposed epidemiological model 
for investigating the behaviour of the contagion spreading 
and Section~\ref{sec:04} shows how the spread of infection occurs in the network. 
Section~\ref{sec:05} details the optimal control problem and discusses 
policy implications. Finally, Section~\ref{sec:06} summarizes 
and concludes our study.


\section{Data}
\label{sec:02}

The complex study of country interrelations shows which national banking systems 
are most exposed to a particular country, both on an immediate counterparty basis 
and on an ultimate risk basis. Claims on an immediate counterparty basis capture 
lending to a borrower that resides in the counterparty country, while claims 
on an ultimate risk basis capture lending to a borrower in any country 
that is guaranteed by an entity that resides in the counterparty country \cite{BIS1}.
We use the Bank for International Settlements (BIS) 
Consolidated Banking Statistics database (consolidated foreign 
claims of reporting banks -- ultimate risks basis) to extract data from Table~B4
``Residence of counterparty, by nationality of reporting bank'' of \cite{BIS2}.
Table~B4 of \cite{BIS2} provides the total claims on an ultimate risk basis by nationality 
of reporting bank. The country of ultimate risk or where the final 
risk lies is defined as the country in which the guarantor of a financial 
claim resides and/or the country in which the head office of a legally 
dependent branch is located \cite{BIS3}.

For our research, we have chosen 13 (thirteen) European and Non-European developed 
countries (Portugal, Italy, Spain, Ireland, Japan, Belgium, France, 
United Kingdom, Austria, United States, Sweden, Canada, Germany), 
based on statistical data from BIS for the end of 2017.


\section{Epidemiological model}
\label{sec:03}

There are many mathematical models (SIS, SIR, SEIR, MSIR, \ldots) that are widely 
used for modelling infectious diseases \cite{MR3810766,MR3602689}. These 
epidemiological models are based on dividing the population into 
compartments, with the assumption that every individual in the same 
compartment has the same characteristics \cite{MR3844698,MR3815138,MR3703345}.
In this paper, we use the classic Kermack and McKendrick SIR model \cite{Kermack} 
to describe the epidemic spreading. This model is a three-state model 
that describes the proportion of a population that is infected, susceptible 
to infection and recovered after a disease; and assumes that individuals 
that leave one class must enter another \cite{Haran}.
We assume that contagion can be transmitted from a country with an infected 
financial system to another one, which has not yet been infected and, 
after recovery, the country produces immunity for a long time.

Mathematically, the SIR model is described by the following system 
of ordinary differential equations that govern its laws of motion:
\begin{equation}
\label{eqSIR}
\begin{cases}
\displaystyle \frac{dS(t)}{dt} = - \beta S(t) I(t),\\[0.3cm]
\displaystyle \frac{dI(t)}{dt} = \beta S(t) I(t) - \gamma I(t),\\[0.3cm]
\displaystyle \frac{dR(t)}{dt} = \gamma I(t),
\end{cases}
\end{equation}
$t \in [0,T]$, subject to the initial conditions
\begin{equation}
\label{eqIC}
S(0) = S_0 > 0, \quad I(0) = I_0 > 0, \quad R(0) = R_0 \geq 0.
\end{equation}
The total population size $N = S(t)+I(t)+R(t)$ keeps constant for $t\in [0, T]$.
Susceptible individuals ($S$) can become infected with the disease, 
and then move into the infected class. Infected individuals ($I$) 
infect susceptible individuals and continue to exist in the infected 
class before moving into the recovered class. Recovered individuals ($R$) 
are no longer infectious and have immunity \cite{Haran}.

Parameters $\beta$ and $\gamma$ are, respectively, the transmission rate, 
that is, the probability of the spread of infection, 
and the recovery rate.

Imposing economic sense to the epidemiological model, 
we consider the following assumptions:
\begin{itemize}
\item initially, only one country is contagious and no one is recovered:
$S(0)=12$, $I(0)=1$, and $R(0)=0$;

\item a contagious country $I$ can infect a susceptible country $S$ 
through their interconnections if the last one has not enough money 
in reserve to cover possible risk losses;

\item an infected counterparty country has all its foreign claims infected;

\item the possibility of infection is higher when there are
more outstanding debts in the total amount of debts;

\item the possibility that a country performs its obligations as a borrower 
depends on the country's credit rating since it takes into account 
not only countries' debt but also assets.
\end{itemize}

\begin{figure}[ht!]
\centering
\includegraphics[scale=0.5]{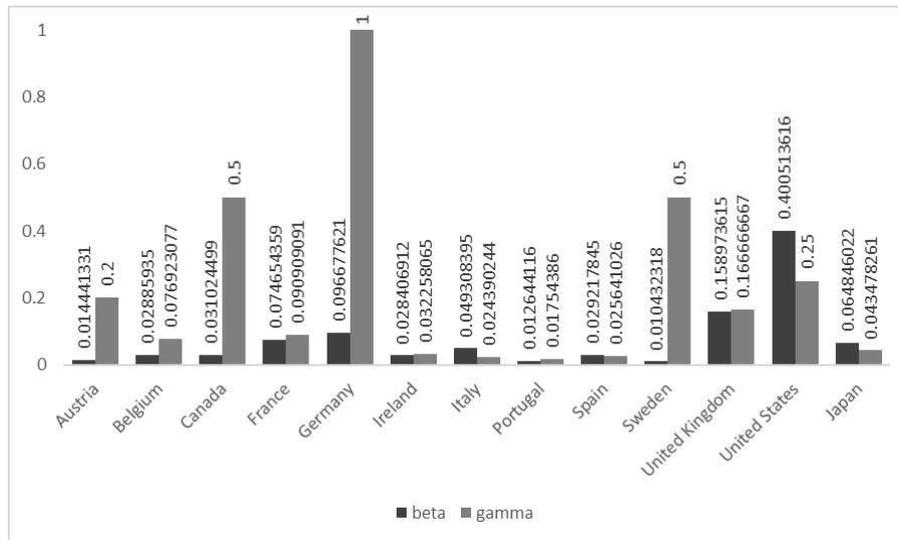}
\vspace*{0.3cm}
\caption{Summary statistics of $\beta$ and $\gamma$ parameters 
for the 13 European and Non-European developed countries considered in our study.} 
\label{fdata}
\end{figure}

Based on the above assumptions, the infection spreading rate 
and the recovery rate are calculated as follows:
\begin{equation}
\label{eqbeta}
\beta_{i} = \frac{\sum\limits_{j=1}^{13} a_{ij}} 
{\sum\limits_{i=1}^{13}\sum\limits_{j=1}^{13} a_{ij}},
\quad i \in \left\{1, \ldots, 13\right\},
\end{equation}
\begin{equation}
\label{eqgamma}
\gamma_{i} = \frac{1} {101-C_{i}}, 
\quad i \in \left\{1, \ldots, 13\right\}, 
\end{equation}
where the exposure $a_{ij}$ is calculated using the data available from BIS 
and the credit rating $C_{i}$ is taken from Trading Economics \cite{Rating}. 
The computed values of contagion spreading rate and the speed 
of recovery are given in Figure~\ref{fdata}.

The model \eqref{eqSIR}--\eqref{eqIC} describes that the contagion begins with 
a country that spreads a disease by infecting its neighbours, where the number 
of new infectious countries generated by the initial one is called the reproduction rate. 
The reproduction rate $\mathcal{R}_0 = \displaystyle \frac{\beta}{\gamma}$ 
is greater than $1$ when an infected country, on average, 
spreads it to at least one additional country, that will 
then do the same thing, and so on. It leads to an epidemic, with the number 
of countries, that receive contagion, growing exponentially.
In contrast, contagion is generally considered to fail when $\mathcal{R}_0$ is less than $1$. 

By solving the initial value problem \eqref{eqSIR}--\eqref{eqIC}, 
we obtain results according to which the epidemiological model shows 
a different behaviour of contamination, depending 
on the values of the $\beta$ and $\gamma$ parameters
of the first originally infected country.

\begin{figure}[htbp]
\centering{
\includegraphics[scale=.5]{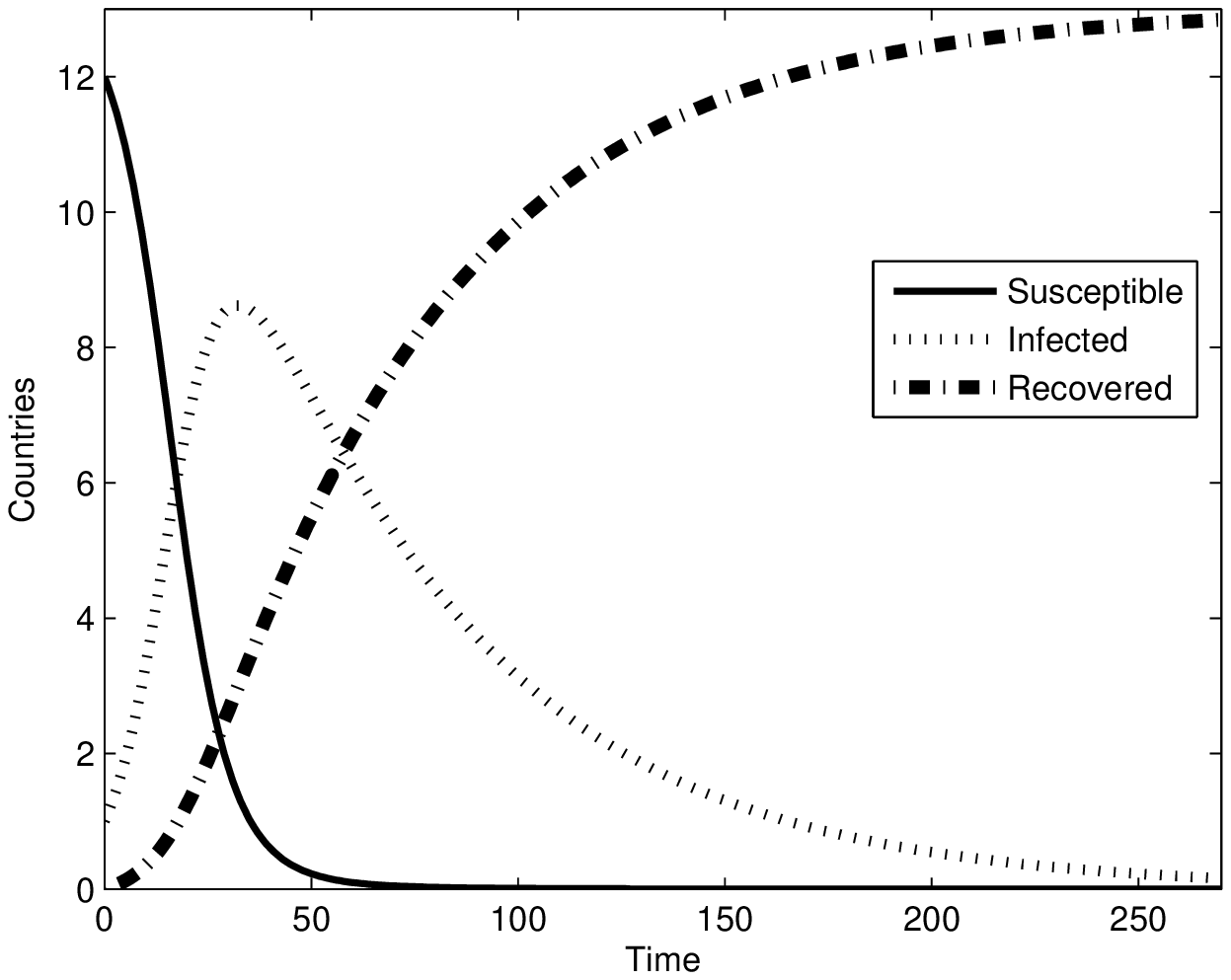}
\includegraphics[scale=.5]{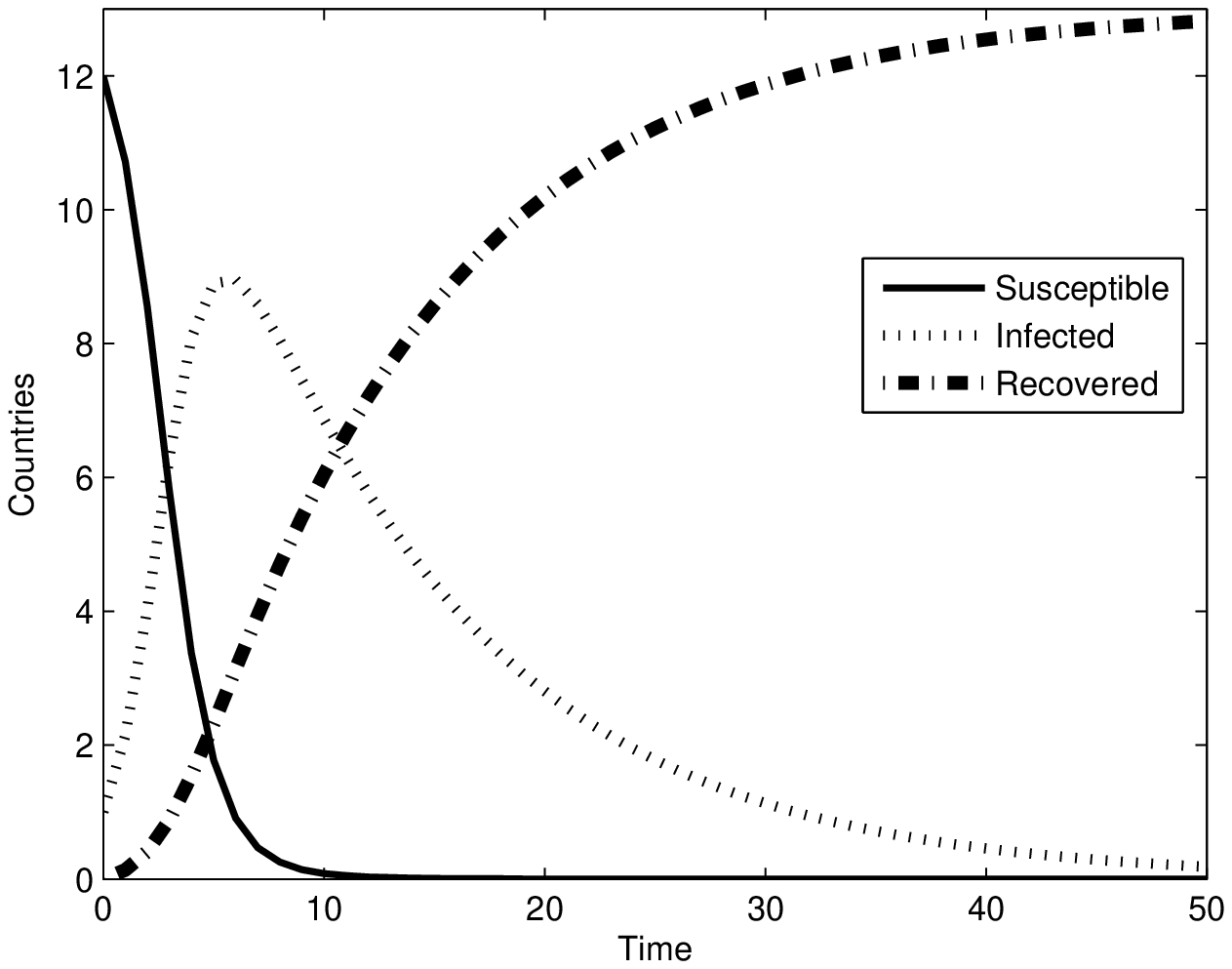}\\
\includegraphics[scale=.5]{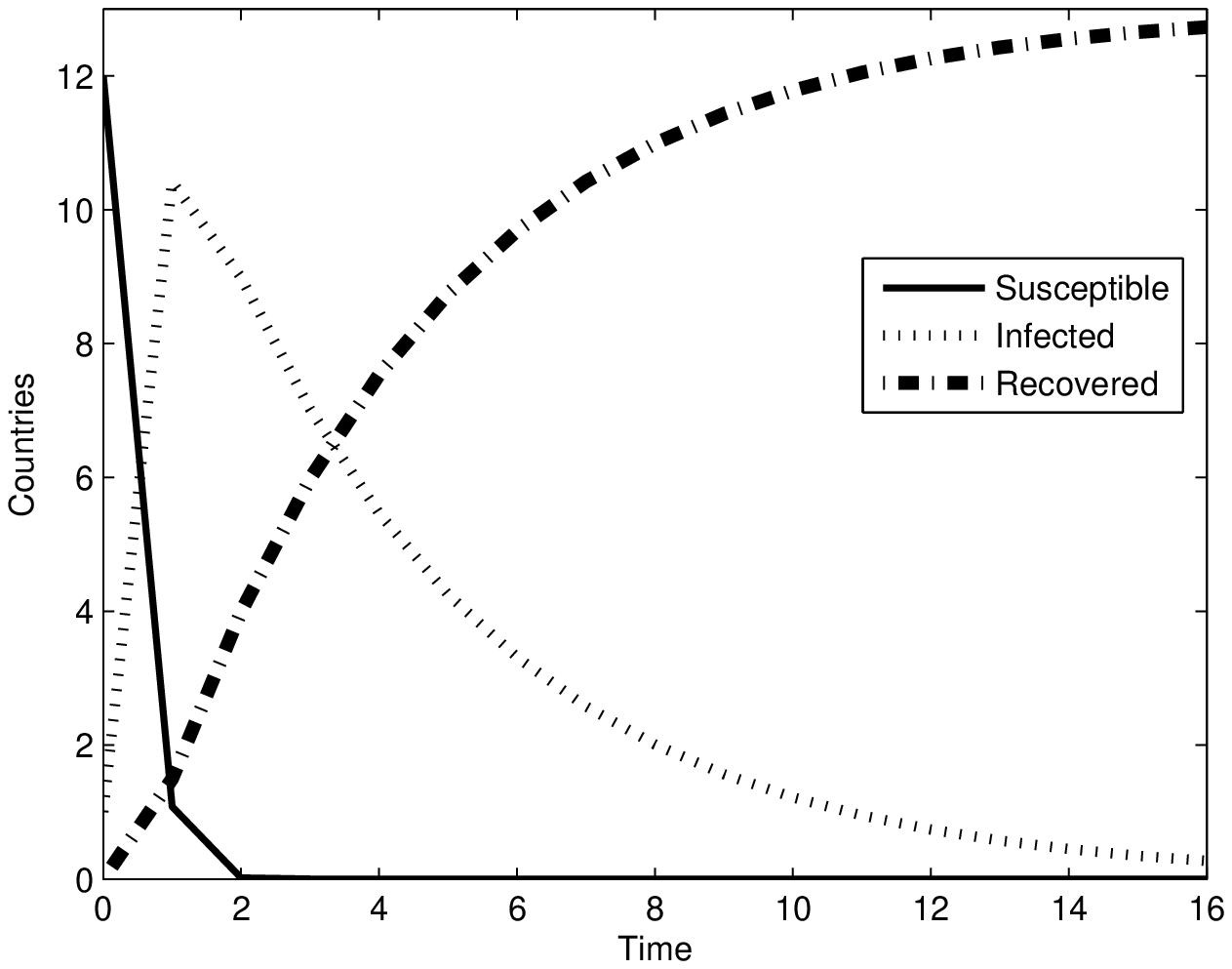}
\includegraphics[scale=.5]{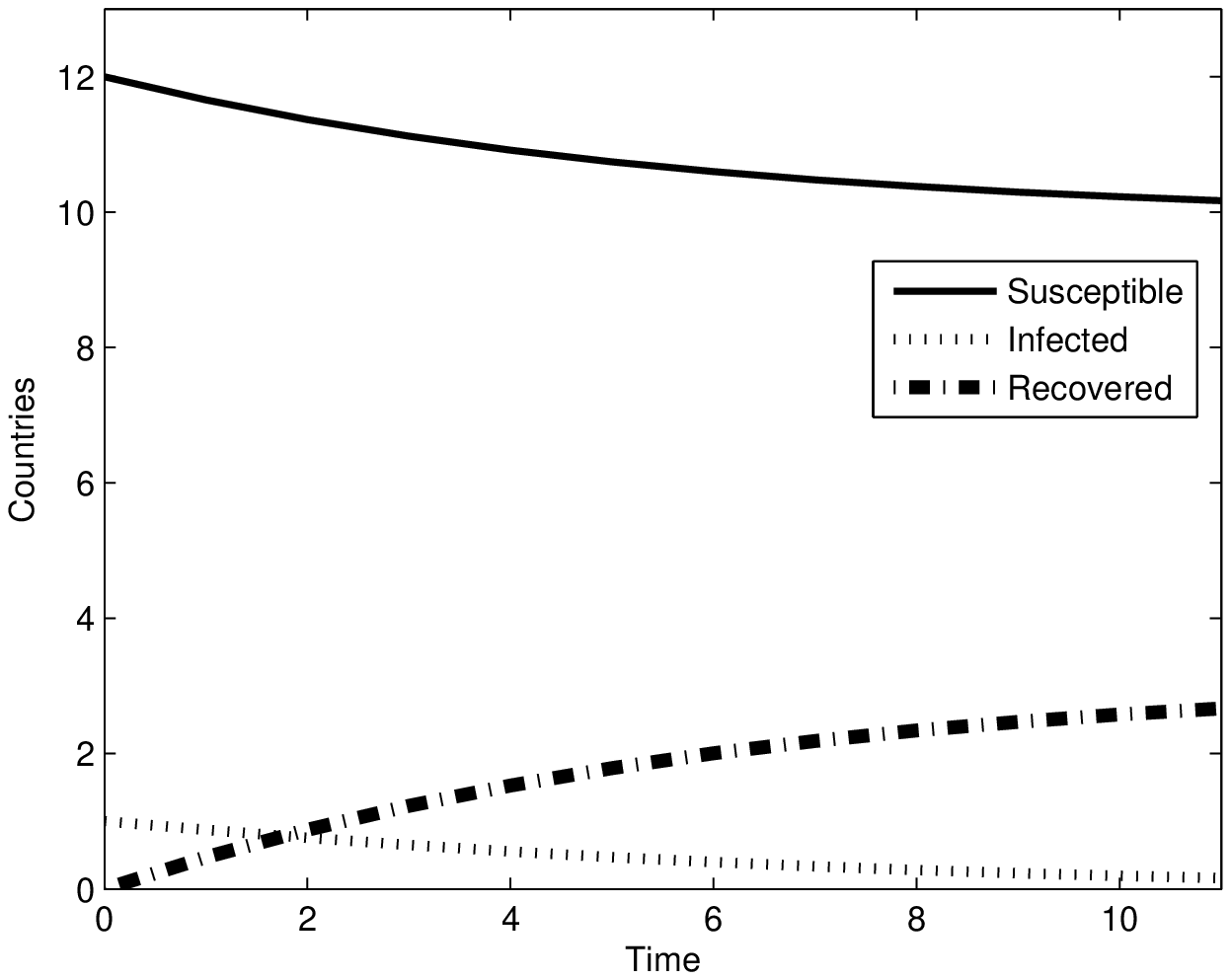}
}
\caption{The SIR contagion risk model. From upper-left to lower-right: 
Portugal, France, USA and Canada as initially infected country.}
\label{SIR}
\end{figure}

All countries in our sample can be divided into 4 groups, depending 
on the value of $\gamma$ and $\mathcal{R}_0$: see Table~\ref{tabular:gamma}.
\begin{table}[ht!]
\doublerulesep 0.1pt
\tabcolsep 7.8mm
\centering
\caption{\rm Grouping of the 13 considered countries, depending on the $\gamma$ parameter and $\mathcal{R}_0$.}
\label{tabular:gamma}
\vspace*{2mm}
\renewcommand{\arraystretch}{1.3}
\setlength{\tabcolsep}{22pt}
\begin{center}
\footnotesize{
\begin{tabular*}{12 cm}{cccc} 
\hline
\hline
\hline
\raisebox{-2ex}[0pt][0pt]{Group 1}&
\raisebox{-2ex}[0pt][0pt]{Group 2}&
\raisebox{-2ex}[0pt][0pt]{Group 3}&
\raisebox{-2ex}[0pt][0pt]{Group 4}\\
\raisebox{-2ex}[0pt][0pt]{$\gamma < 0.5$} &
\raisebox{-2ex}[0pt][0pt]{$0.5 \leq \gamma < 0.1$} &
\raisebox{-2ex}[0pt][0pt]{$0.1 < \gamma < 1$}&
\raisebox{-2ex}[0pt][0pt]{$0.1 < \gamma \leq 1$}\\
\raisebox{-2ex}[0pt][0pt]{}&
\raisebox{-2ex}[0pt][0pt]{}&
\raisebox{-2ex}[0pt][0pt]{$\mathcal{R}_0 > 0.9$}&
\raisebox{-2ex}[0pt][0pt]{$\mathcal{R}_0 < 0.9$}\\
\\ \hline
PT & FR & USA & CA\\
IT & BE & GB  & AT\\
ES &    &     & SE\\
IE & 	&     & DE\\
JP &    &     &   \\ \hline\hline\hline
\end{tabular*}
}
\end{center}
\renewcommand{\arraystretch}{1}
\end{table}
For illustration purposes, we present four different scenarios of contagion spreading, 
depending on the group the initially infected country belongs: see Figure~\ref{SIR}.

The first scenario is when the infection begins in Portugal. 
The results demonstrate that in the case where Portugal is the 
starting point for the spread of financial contamination, 
this process takes a long time, the infection reaches its peak 
in about 3 years, and the process of recovery also occurs very slowly.

The second scenario is when the initially infected country is France. 
Having a higher risk of infection than Portugal, but significantly higher 
credit rating than the Portuguese financial system, the infection
reaches a contagion-free equilibrium ($I(T)=0$) five times faster
than in previous scenario.

For the third scenario, we chose the United States of America (USA).
As for the USA, the $\beta$ parameter is higher than $\gamma$, that is,
the contamination rate is higher than recovery. This translates
in the shortest period of time the largest number of countries become infected.

Our fourth scenario has Canada as the initially infected country.
The situation with Canada is the most optimistic.
Indeed, the system reaches its equilibrium in a slightly shorter period 
of time than with the USA and, simultaneously, there is 
a minimum number of countries affected because $\mathcal{R}_0 < 1$.

In next sections of network analysis (Section~\ref{sec:04})
and optimal control (Section~\ref{sec:05}), we consider
only the problematic situations, that is, the cases 
when $\mathcal{R}_0 > 1$: the results of simulations 
for contagion spreading from Canada are not shown because 
in this situation there is no need to take control measures.


\section{Network}
\label{sec:04}
 
To visualize the spread of infection among participants in our sample, 
a network analysis was carried out. Countries, according to graph theory, 
are represented in the form of nodes (vertices) while edges (links) 
mean the existence of financial ties between the countries.
According to the data from BIS (Table~B4 of \cite{BIS2}),
all $n = 13$ countries that were chosen for our sample are connected to each other, 
but none with themselves. In other words, each vertex has the same number of neighbours
and the network of countries' interrelations, for our sample, can be represented 
as a complete graph $K_n$, where each vertex has the same degree $n-1$ 
and the graph has $n(n-1)/2$ edges. According to these characteristics, 
the network topology in our research was chosen as a fully connected network.

\begin{figure}[htbp]
\centering{
{\label{Netpt0}\includegraphics[scale=.215]{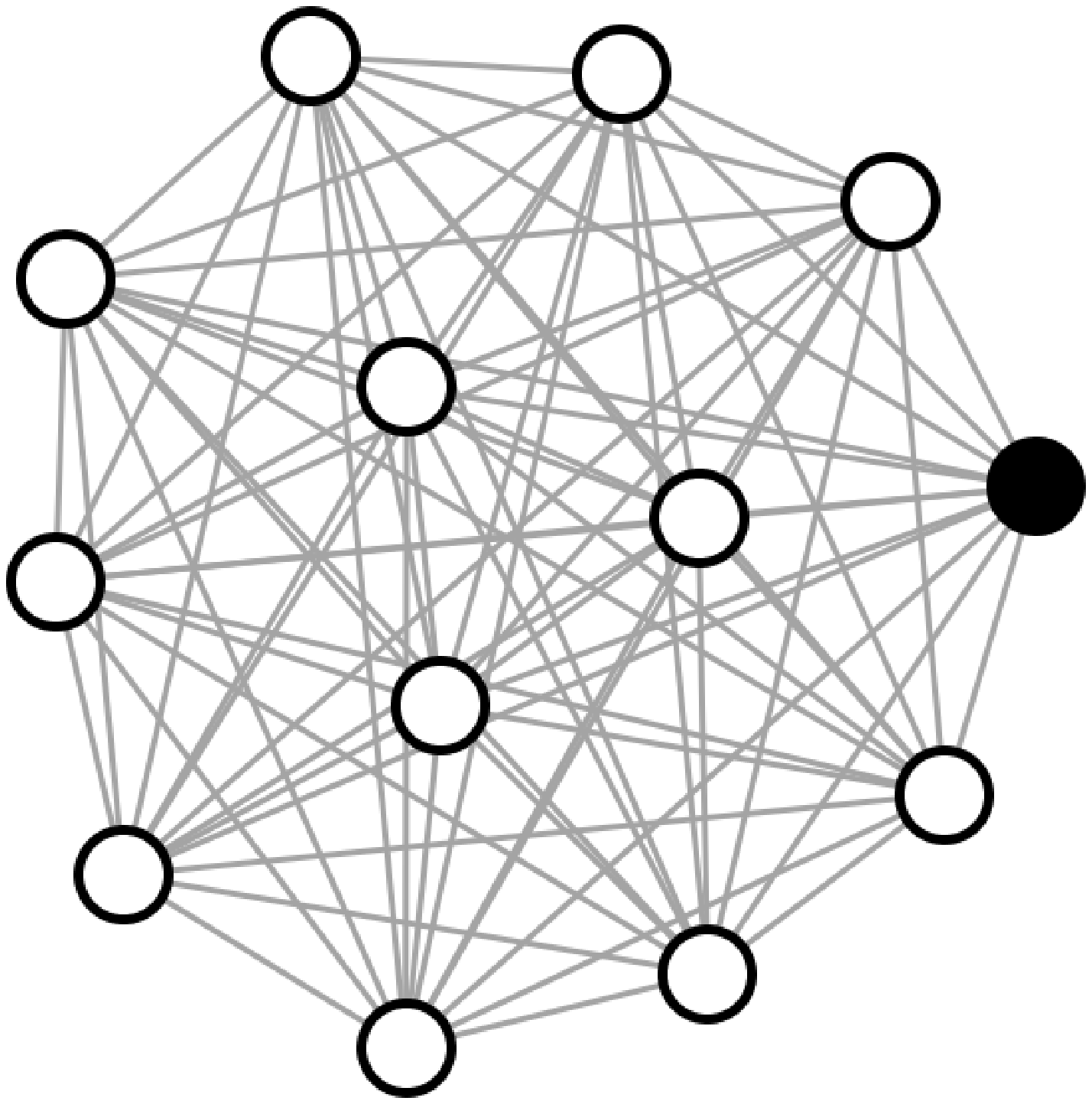}}
{\label{Netpt35}\includegraphics[scale=.215]{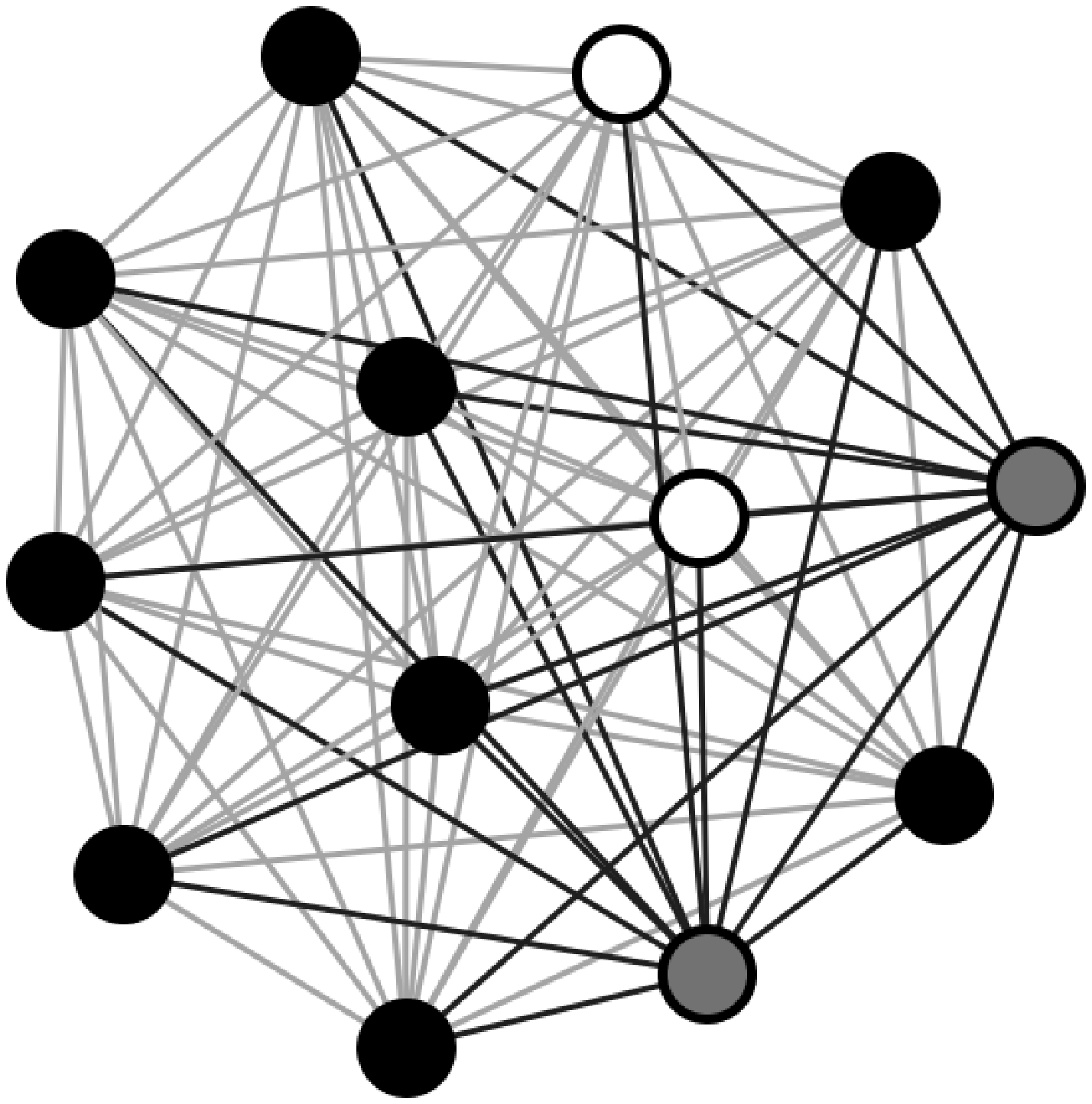}}
{\label{Netpt162}\includegraphics[scale=.215]{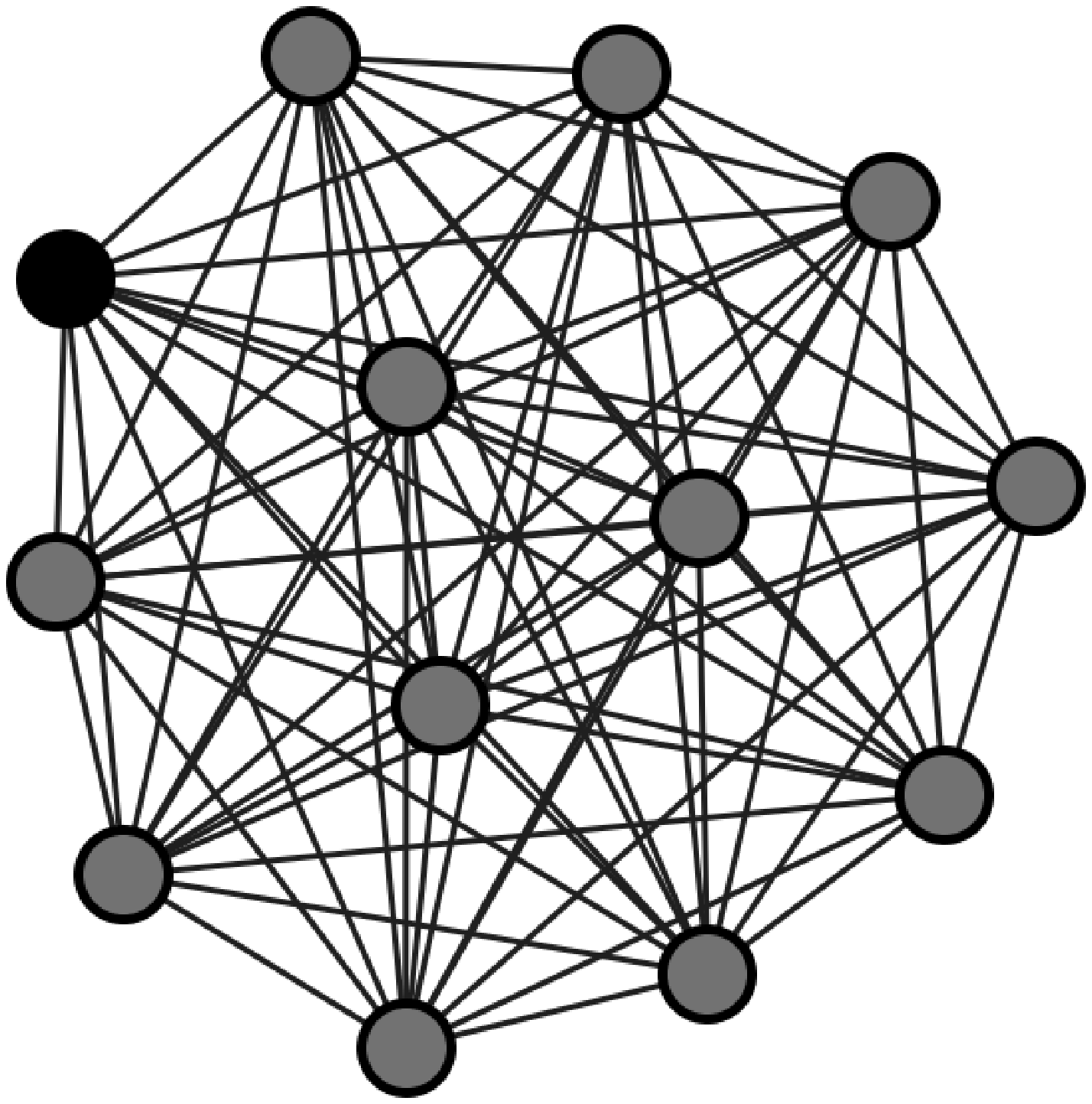}}
{\label{Netpt270}\includegraphics[scale=.215]{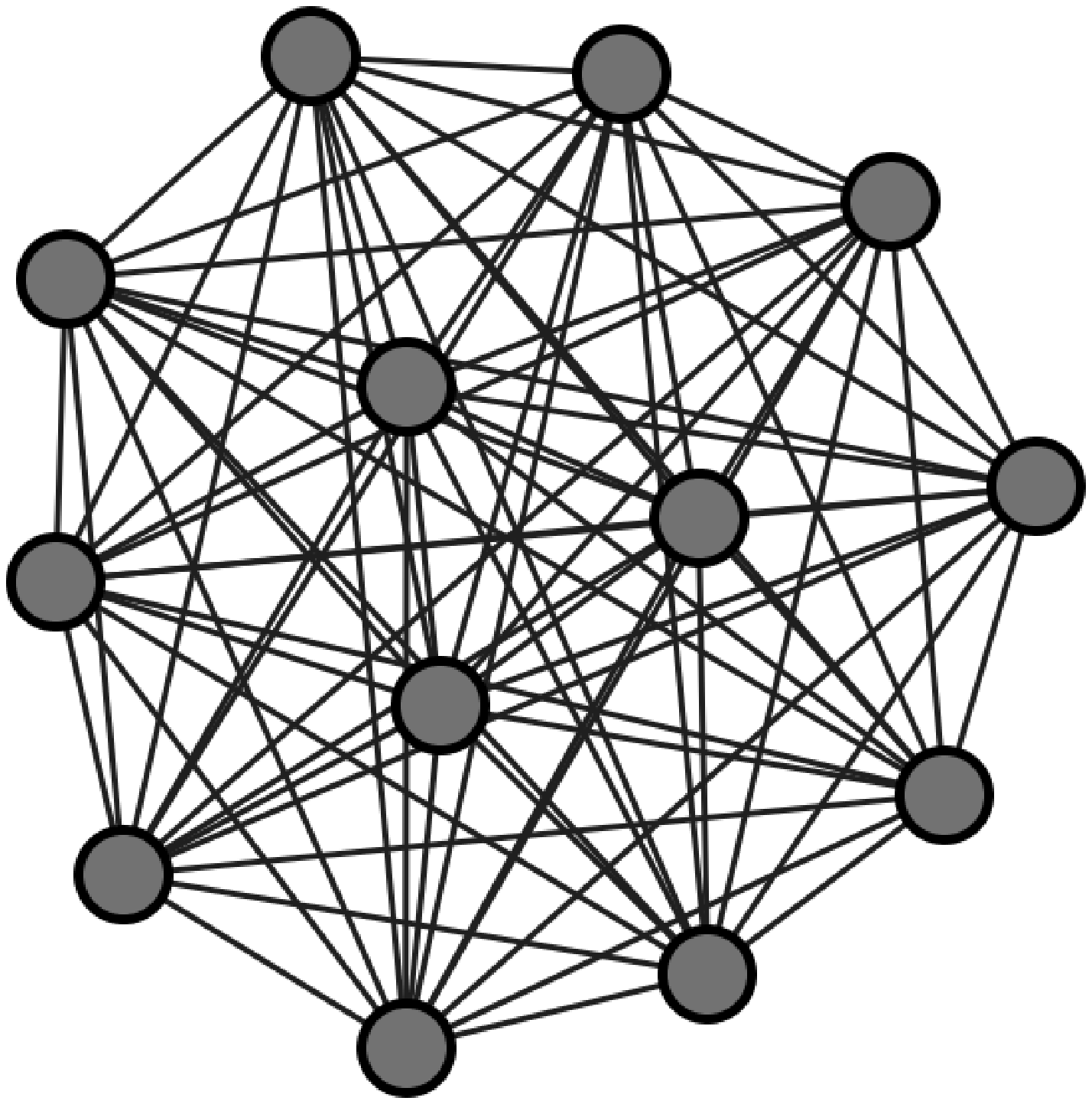}}\\
{\label{Netfr0}\includegraphics[scale=.215]{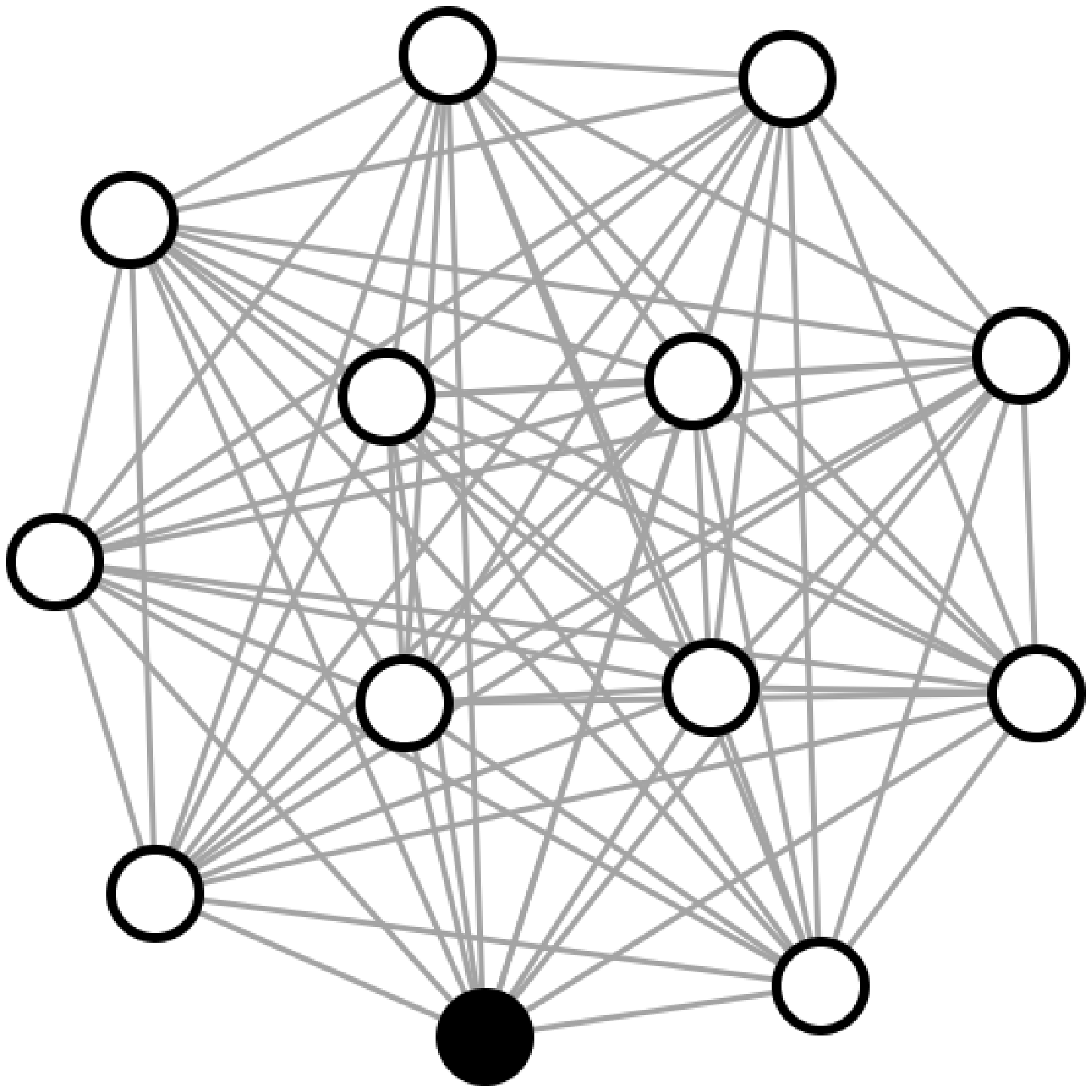}}
{\label{Netfr5}\includegraphics[scale=.215]{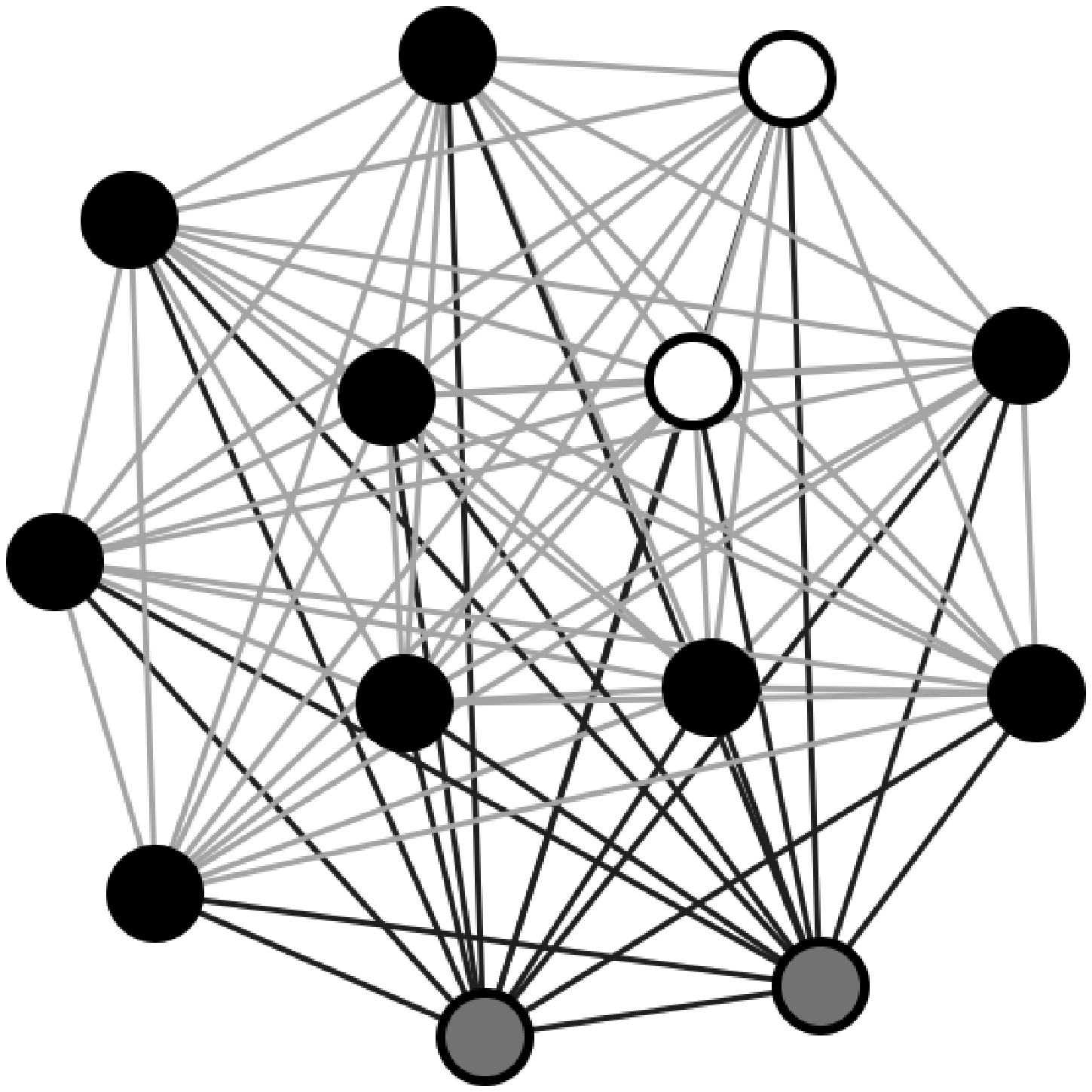}}
{\label{Netfr49}\includegraphics[scale=.215]{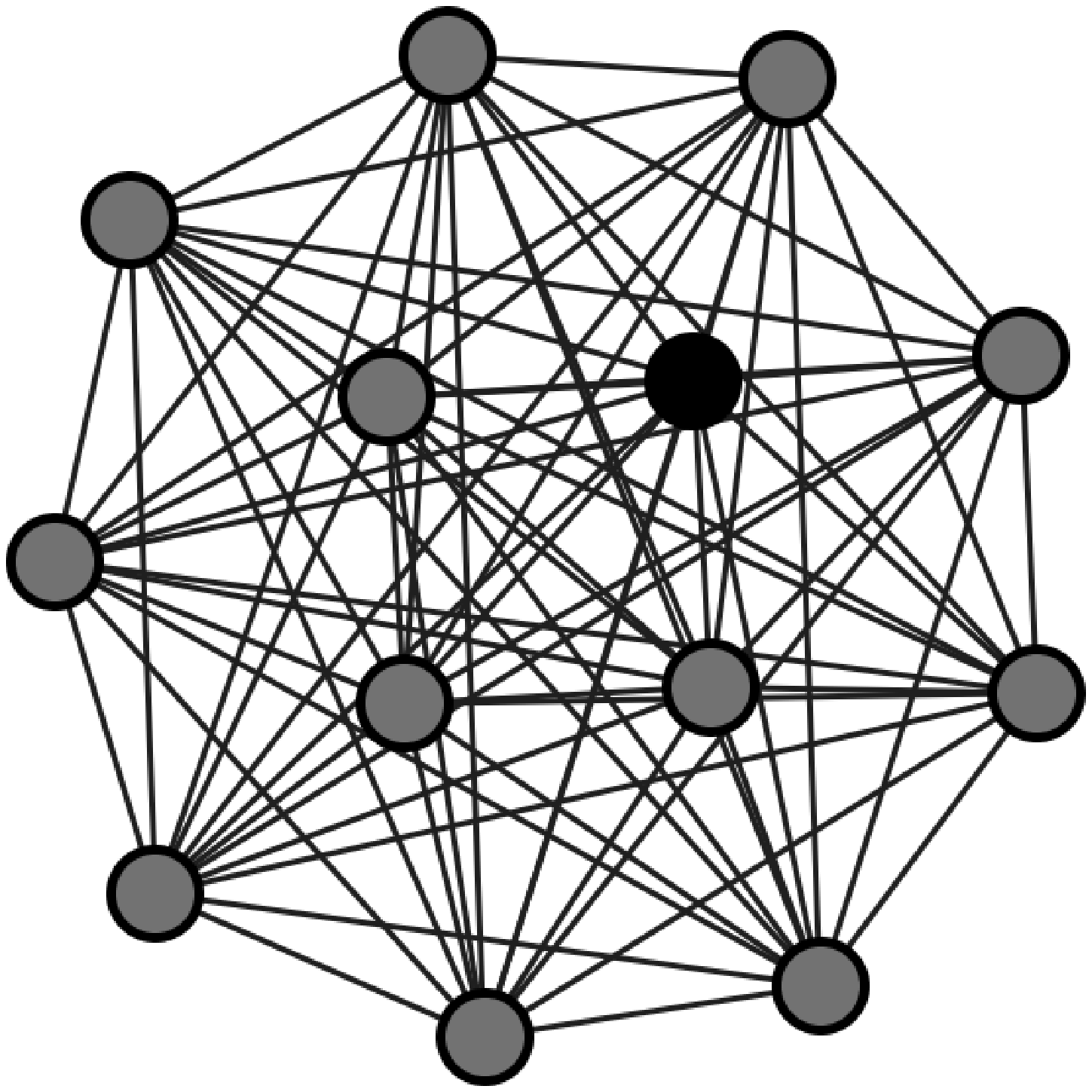}}
{\label{Netfr50}\includegraphics[scale=.215]{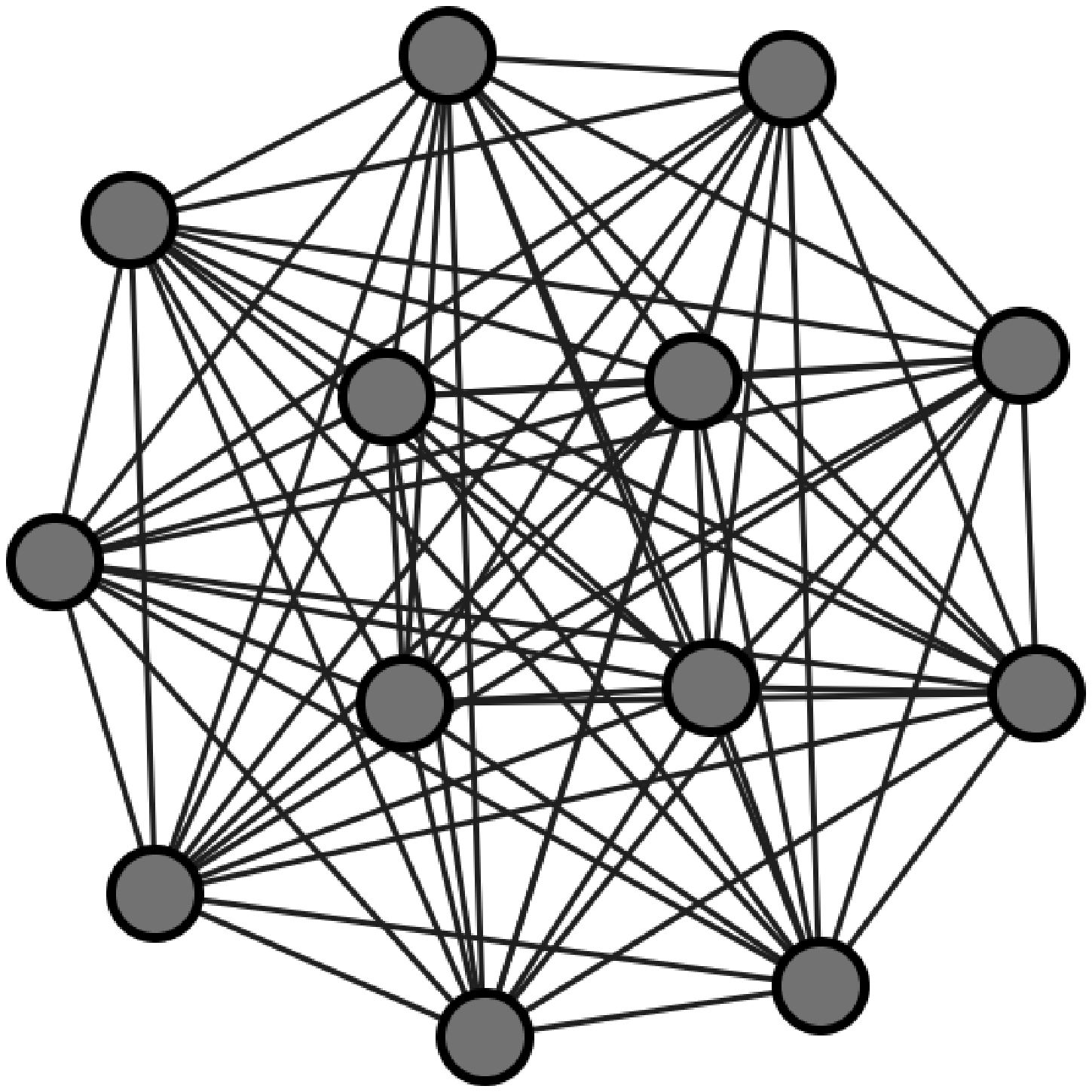}}\\
{\label{Netusa0}\includegraphics[scale=.215]{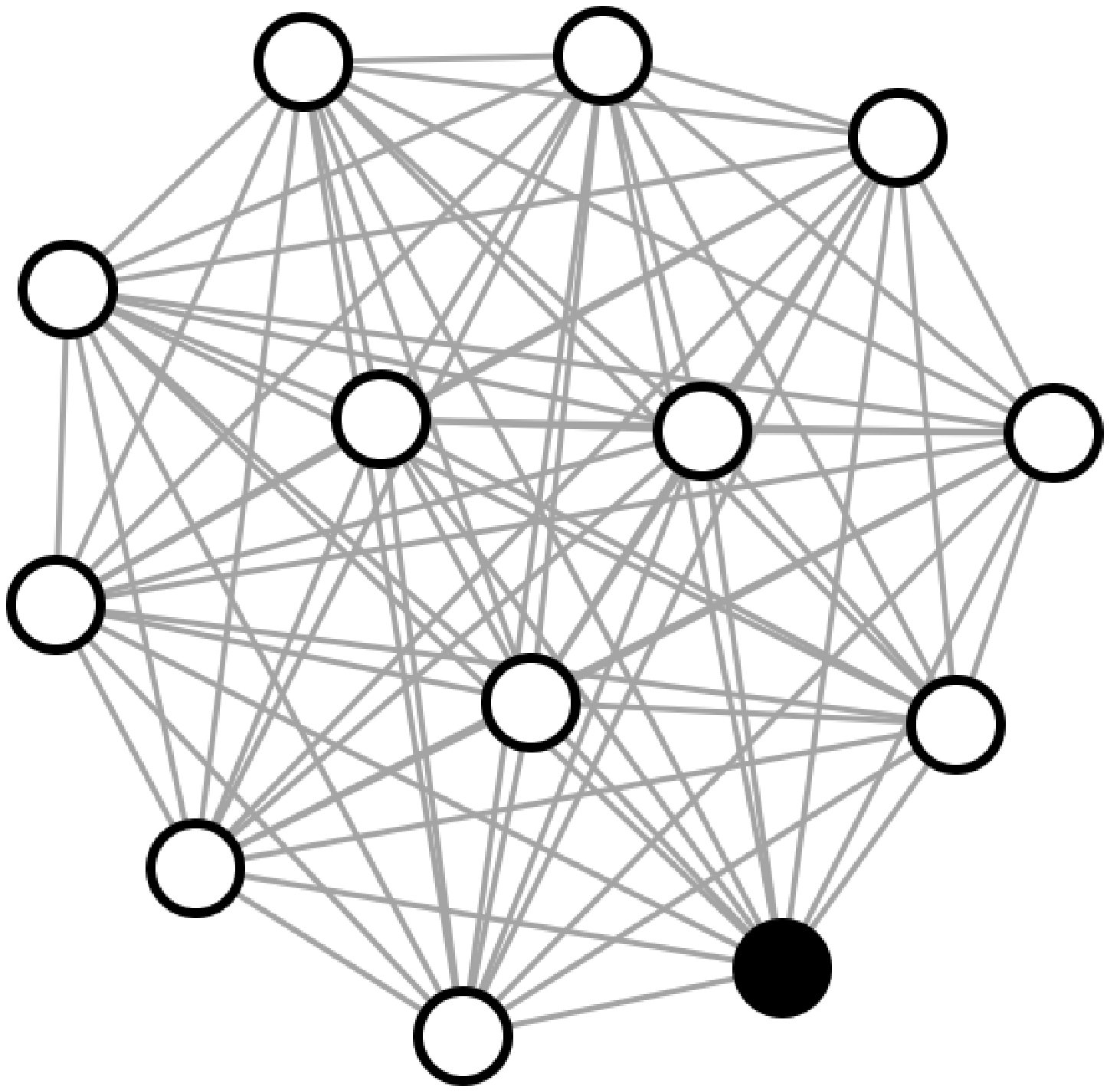}}
{\label{Netusa2}\includegraphics[scale=.215]{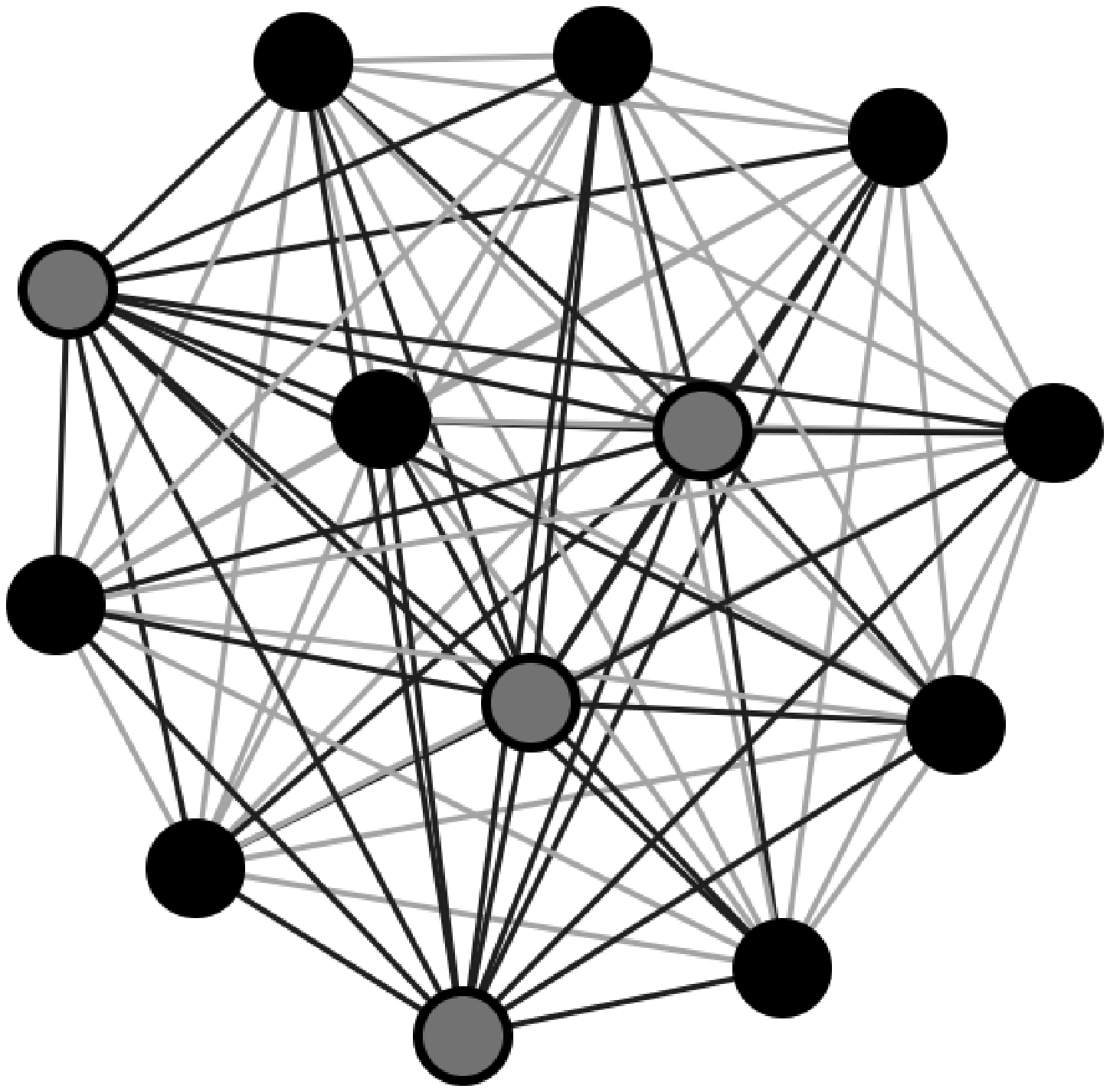}}
{\label{Netusa15}\includegraphics[scale=.215]{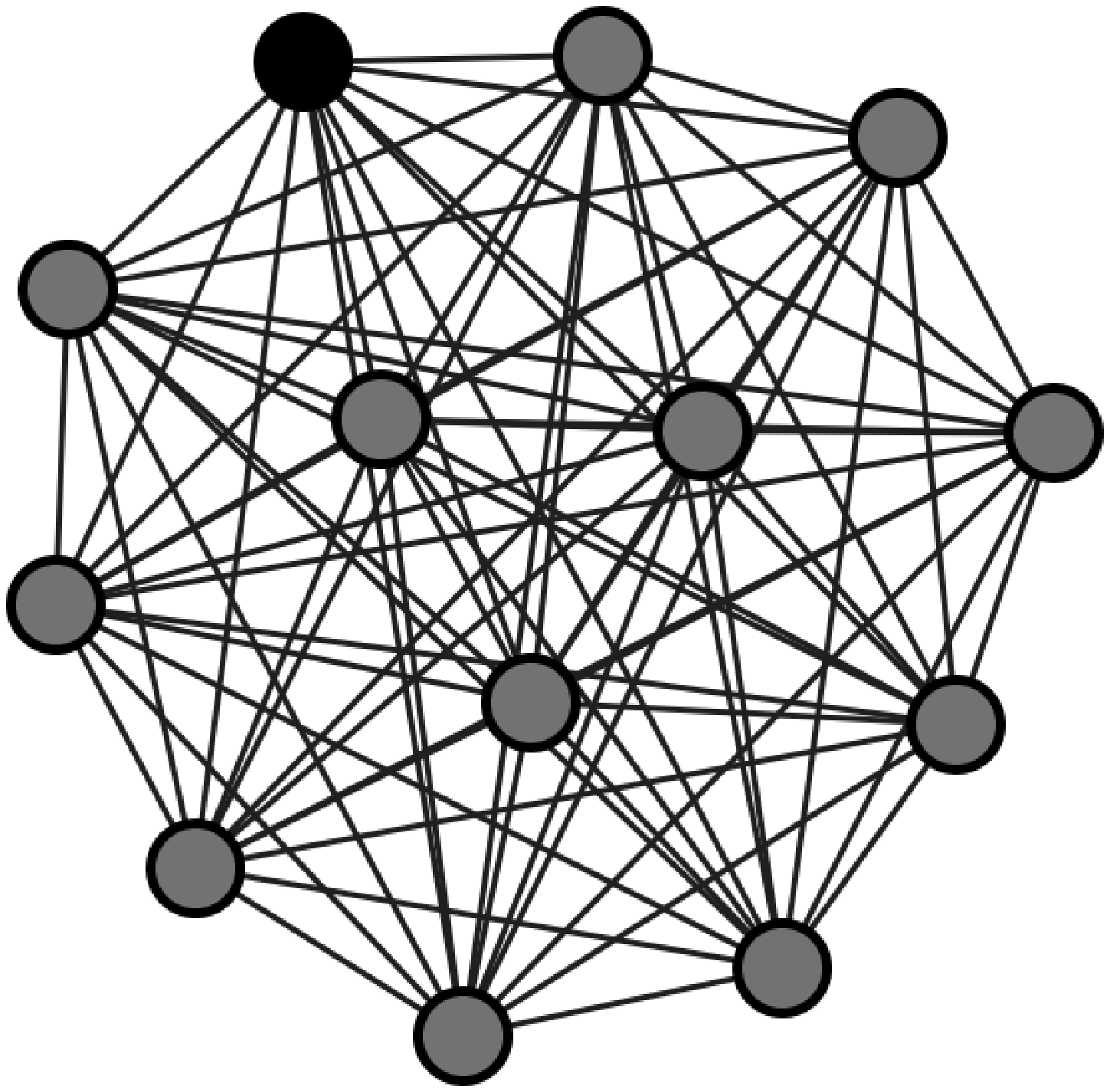}}
{\label{Netusa16}\includegraphics[scale=.215]{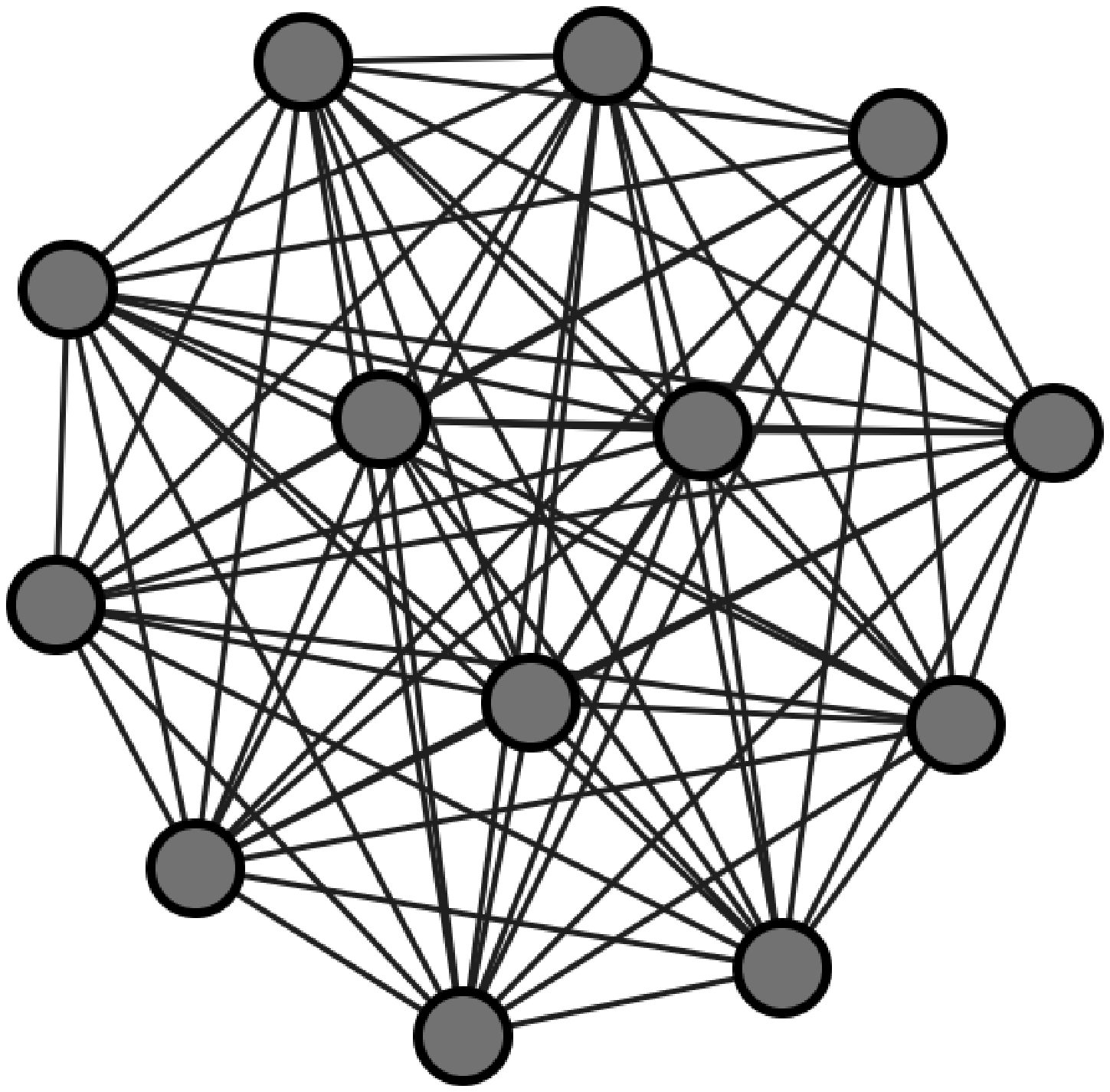}}
}
\caption{Virus spreading in the network of 13 countries with parameters 
$\beta$ and $\gamma$ as in Figure~\ref{fdata}. From upper-row 
to lower-row: Portugal ($T=0$, $T=35$, $T=162$, $T=270$ months), 
France ($T=0$, $T=5$, $T=49$, $T=50$ months) 
and USA ($T=0$, $T=2$, $T=15$, $T=16$ months).}
\label{Net}
\end{figure}

The dynamics of contagion propagation in the network was investigated
using the NetLogo multi-agent programmable modeling environment 
\cite{NetLogo}. The results are shown in Figure~\ref{Net}
for the three endemic scenarios.
 
Each node in the network represents a random country from our sample. 
A white node represents a susceptible country, black an infected country.
The first column from Figure~\ref{Net} represents the initial moment ($T_0$), 
where only one node has an infection, while others are susceptible. 
In each time step, an infected node attempts to infect all of its susceptible 
neighbours. The recovered nodes, represented in grey, cannot be infected. 
Susceptible neighbours will get an infection and change their colour 
to black with probability $\beta$. Infected nodes will be recovered 
and change their colour to grey with probability $\gamma$. 
When a node becomes recovered, the links between it and its neighbours 
change their colour to grey and are no longer possible 
vectors for contagion spreading \cite{netlogprog}.
The results we obtain here are in agreement 
with those of Section~\ref{sec:03} 
(compare Figures~\ref{SIR} and \ref{Net}).


\section{Optimal control}
\label{sec:05}

All financial institutes need to be under financial supervision 
in order to prevent contagion spreading and avoid serious consequences.
Such a role, in each state, is carried out by the Central Bank, 
which acts as a supervisory competent authority. Since one of the main 
goals is to prevent the global spread of infection, we pose 
the following optimal control problem in Bolza form: 
functional
\begin{equation}
\label{eqopt1}
\min {J}[I(\cdot),u(\cdot)]
=I(T)+\int\limits_0^T b u^2(t) dt 
\end{equation}
subject to the control system
\begin{equation}
\label{eqopt2}
\begin{cases}
\displaystyle \frac{dS(t)}{dt} = - \beta S(t) I(t),\\[0.3cm]
\displaystyle \frac{dI(t)}{dt} = \beta S(t) I(t) - \gamma I(t)-u(t)I(t),\\[0.3cm]
\displaystyle \frac{dR(t)}{dt} = \gamma I(t)+u(t)I(t),
\end{cases}
\end{equation}
where $S(0) = S_0$, $I(0) = I_0$, $R(0) = R_0$, $b > 0$, $u(t) \in [0, 1]$.  
The integral in the objective functional \eqref{eqopt1} means a general cost of financial 
assistance, necessary to prevent the spread of contagion and economic decline 
in the period $[0, T]$ \cite{Kostylenko}. 
The control functions $u(\cdot)$ are selected within a special set of admissible controls.
It consists of those functions $u = u(t)$, which are piecewise continuous functions 
on the interval $[0, T]$ with values in $U = [0,1]$.
The control $u(t)$ denotes the rate at which assistance will be provided 
to a contagious country, that is, it is the ratio between the 
financial support from the Central Bank at time $t$ and the financing 
needed by the countries at that time. If the Central Bank provides 
a full support at time $t$, covering all necessary risks, then $u(t) = 1$.
If the country receives no financial lending at $t$, then one has the control $u(t)=0$.
The smaller the value of the optimality criterion $J[I(\cdot),u(\cdot)]$,
the more beneficial the control $u$ is. In our simulations,
the cost of control measures $b$ is taken to be $1.5$, motivated by the value 
of possible recapitalization with state funds considered in \cite{Philippas}. 

The optimal control problem in Bolza form \eqref{eqopt1}--\eqref{eqopt2} 
was rewritten in the following equivalent Mayer form and 
solved in the BOCOP optimal control solver \cite{Bocop}:
\begin{equation}
\label{eqbocop1}
\min {J}[S(\cdot),R(\cdot),Y(\cdot)]
=N-S(T)-R(T)+Y(T)
\end{equation}
subject to
\begin{equation}
\label{eqbocop2}
\begin{cases}
\displaystyle \frac{dS(t)}{dt} = \beta S^2(t)+ \beta S(t)(R(t)-N),\\[0.3cm]
\displaystyle \frac{dR(t)}{dt} = \gamma (N-S(t)-R(t))+u(t)(N-S(t)-R(t)),\\[0.3cm]
\displaystyle \frac{dY(t)}{dt} = bu^2(t),\\[0.3cm]
\displaystyle S(0)=S_0, \quad R(0)=0, \quad Y(0)=0,\\[0.3cm]
u(t) \in [0, 1].
\end{cases}
\end{equation}

\begin{figure}[ht!]
\centering{
\includegraphics[scale=.42]{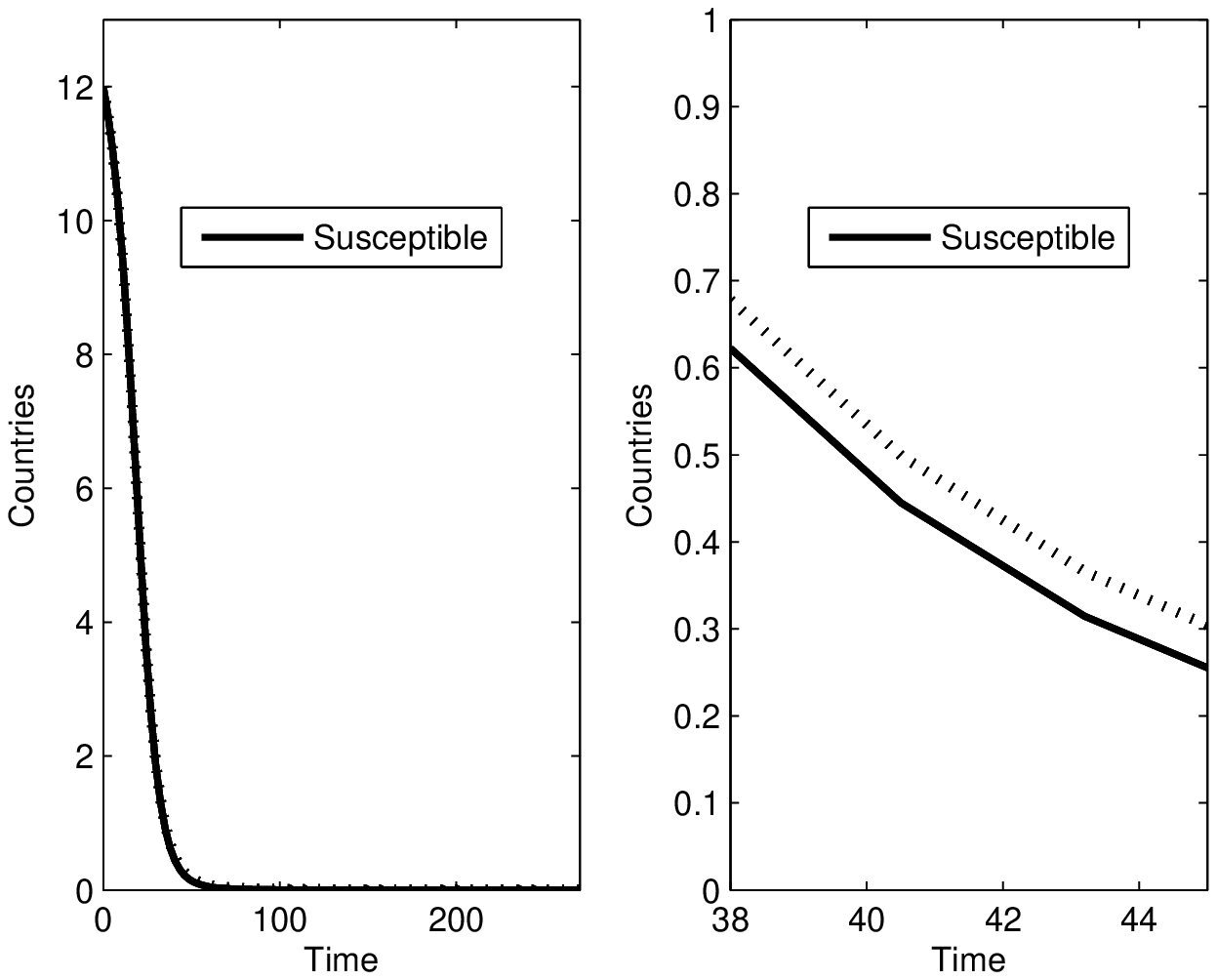}
\includegraphics[scale=.42]{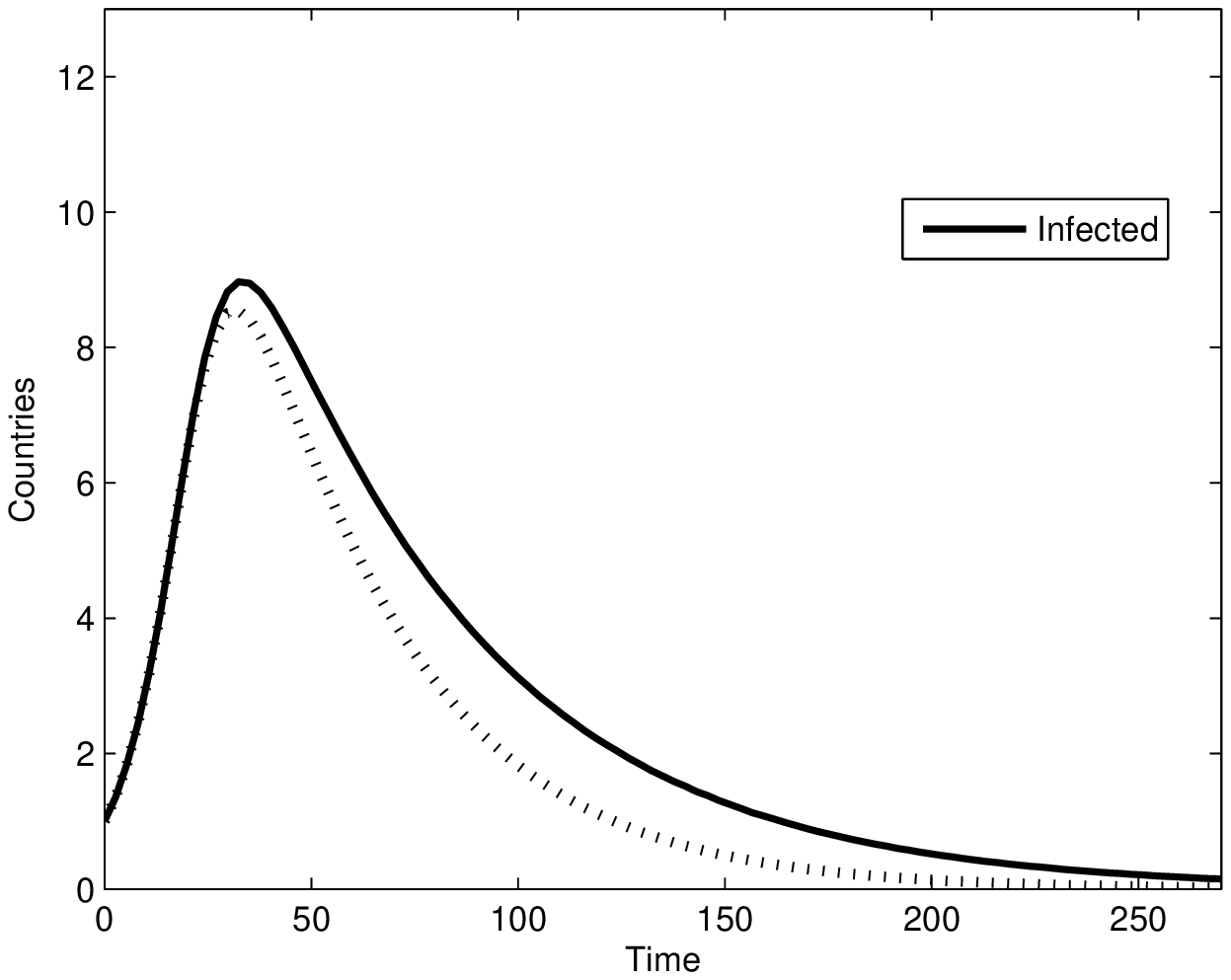}\\
\includegraphics[scale=.42]{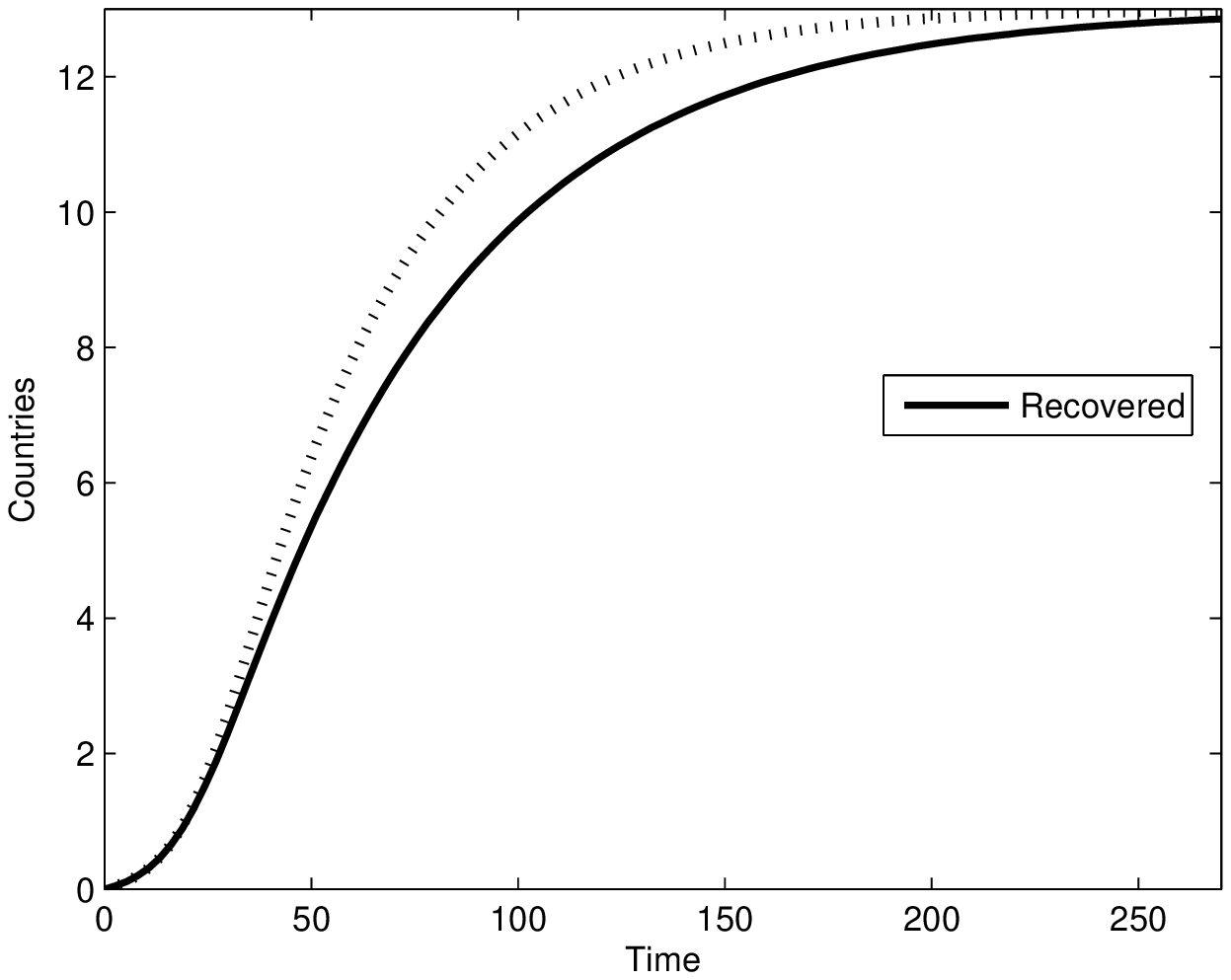}
\includegraphics[scale=.42]{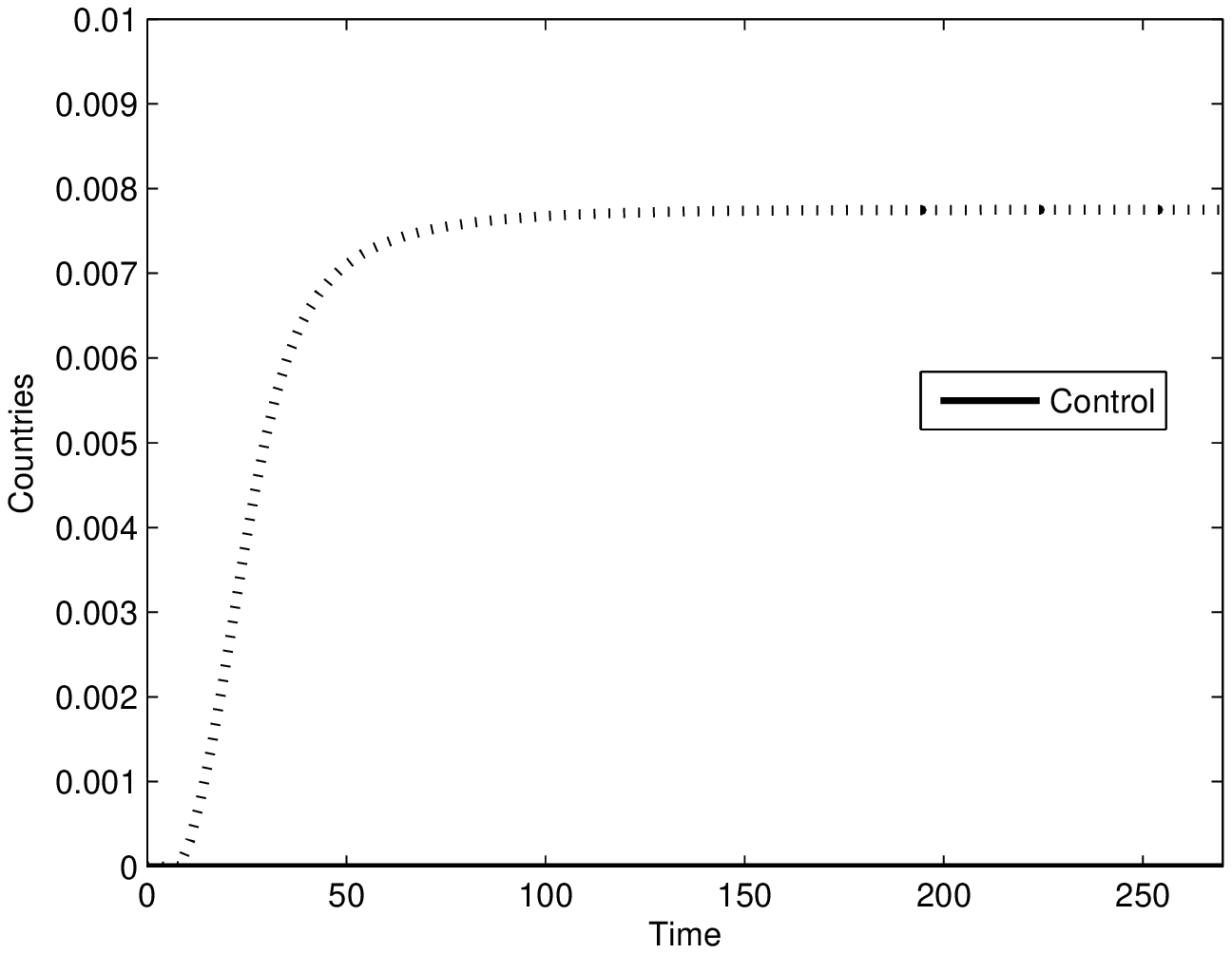}
}
\caption{Contagion risk from Portugal (Scenario 1): 
with optimal control (dotted line) vs without control (continuous line).}
\label{PtControl}
\end{figure}

\begin{figure}[ht!]
\centering{
\includegraphics[scale=.42]{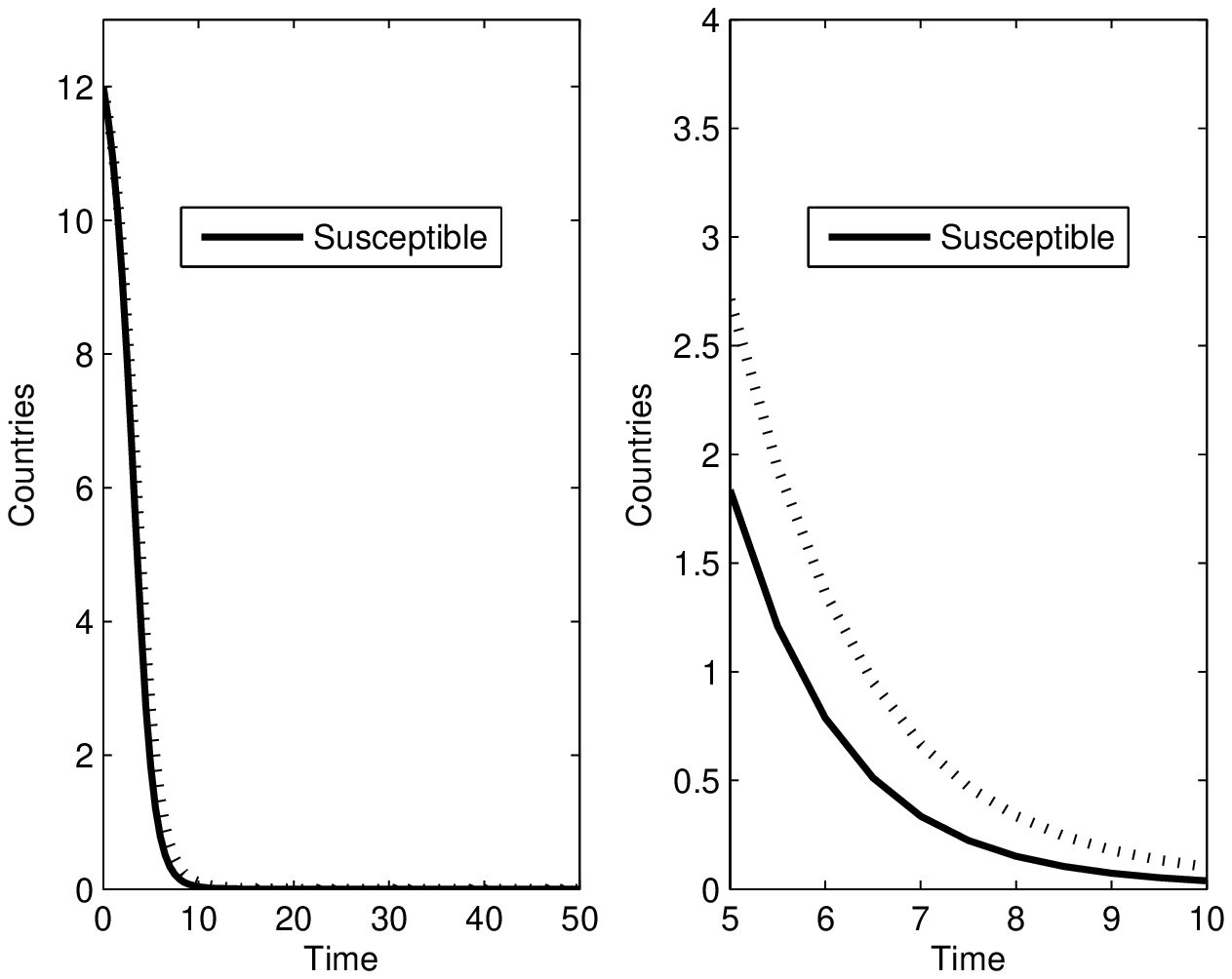}
\includegraphics[scale=.42]{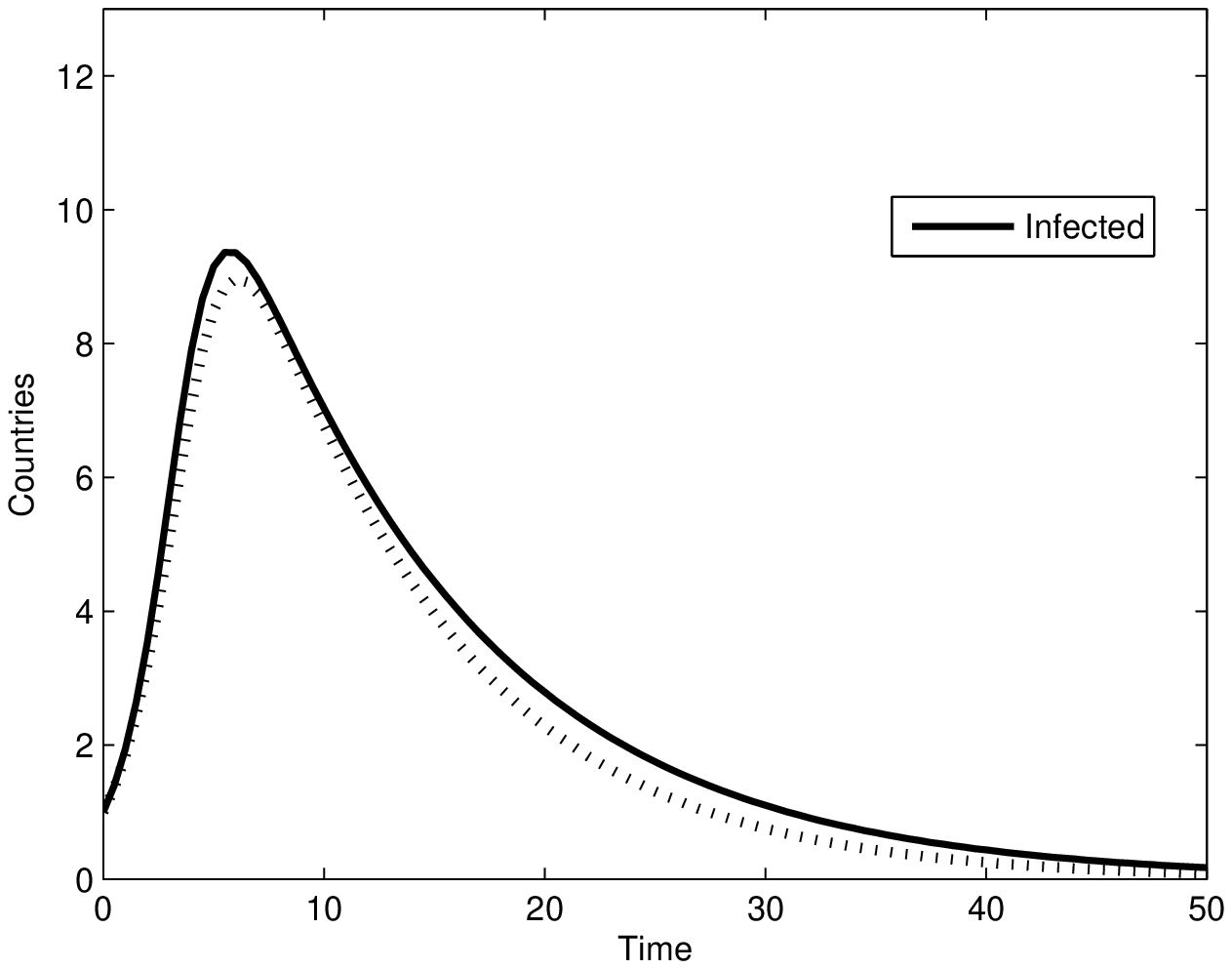}\\
\includegraphics[scale=.42]{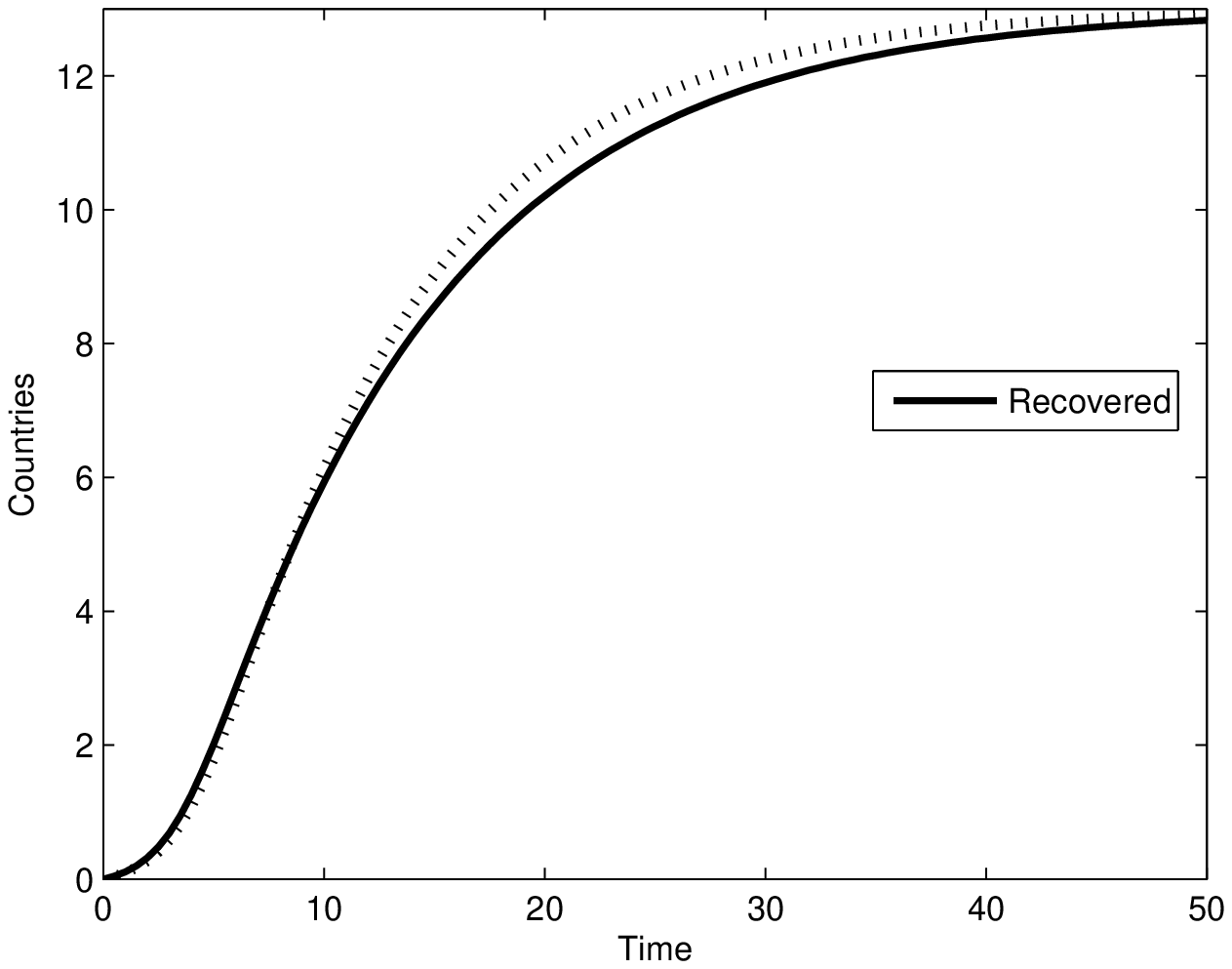}
\includegraphics[scale=.42]{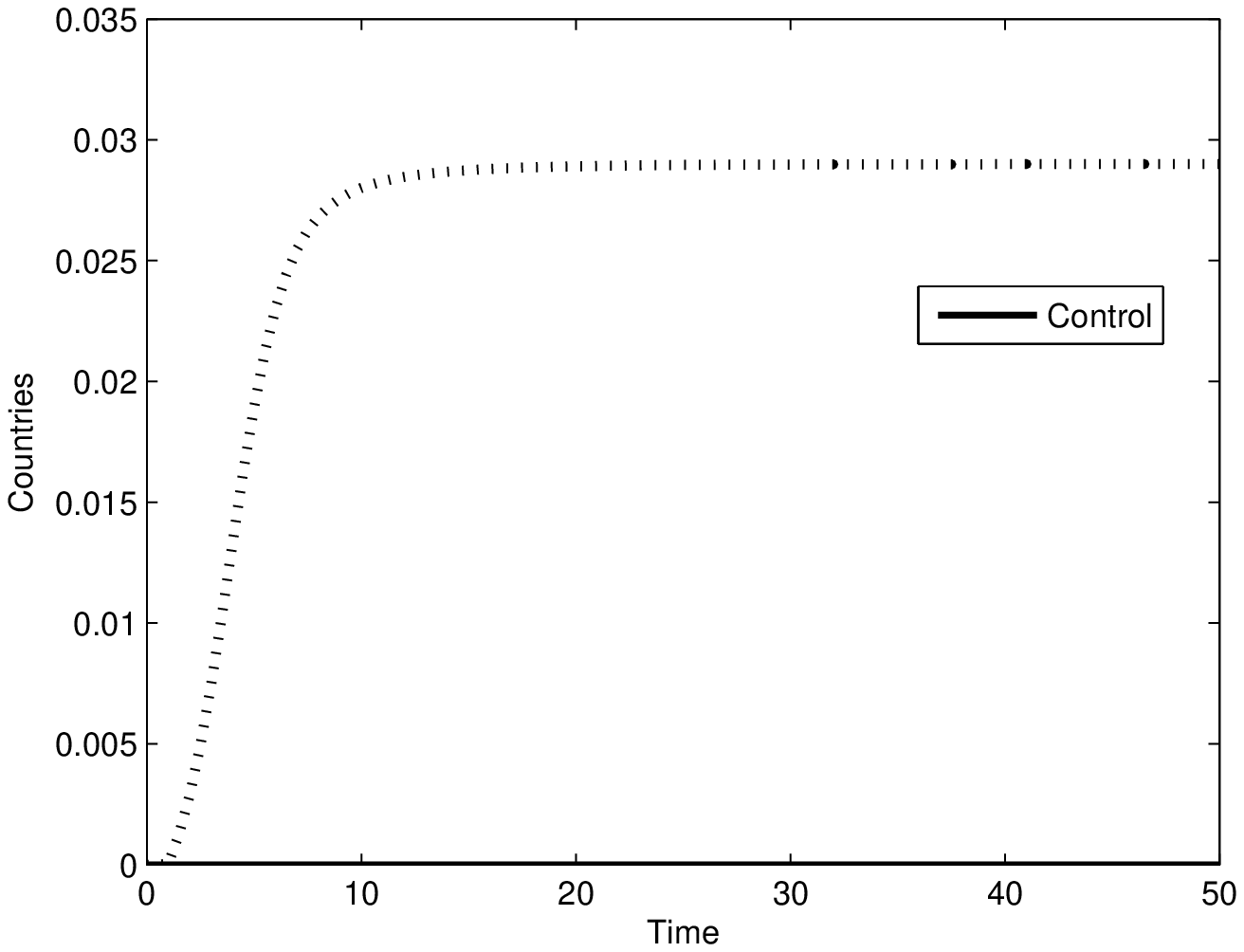}
}
\caption{Contagion risk from France (Scenario 2): with optimal 
control (dotted line) vs without control (continuous line).}
\label{FrControl}
\end{figure}

\begin{figure}[ht!]
\centering{
\includegraphics[scale=.42]{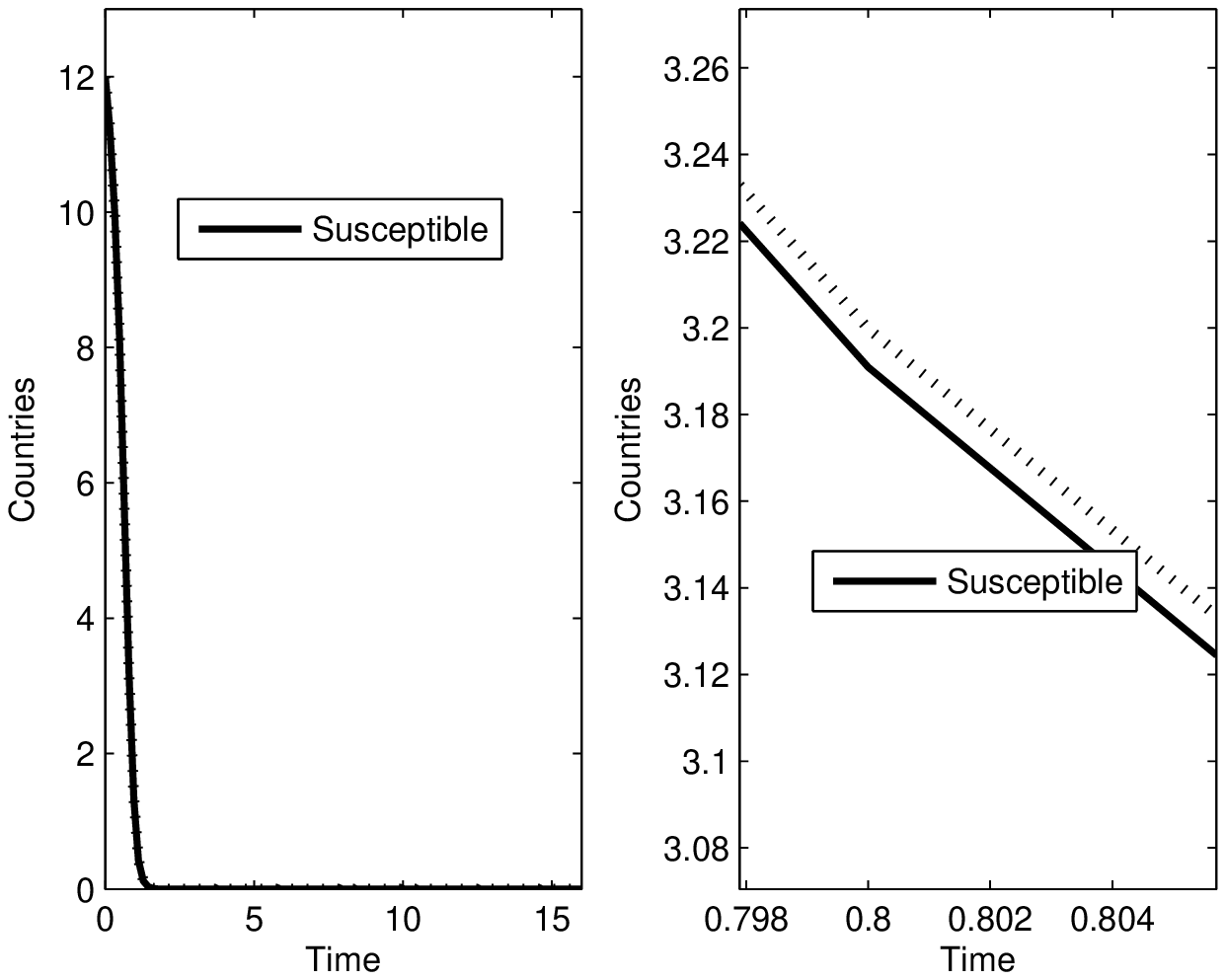}
\includegraphics[scale=.42]{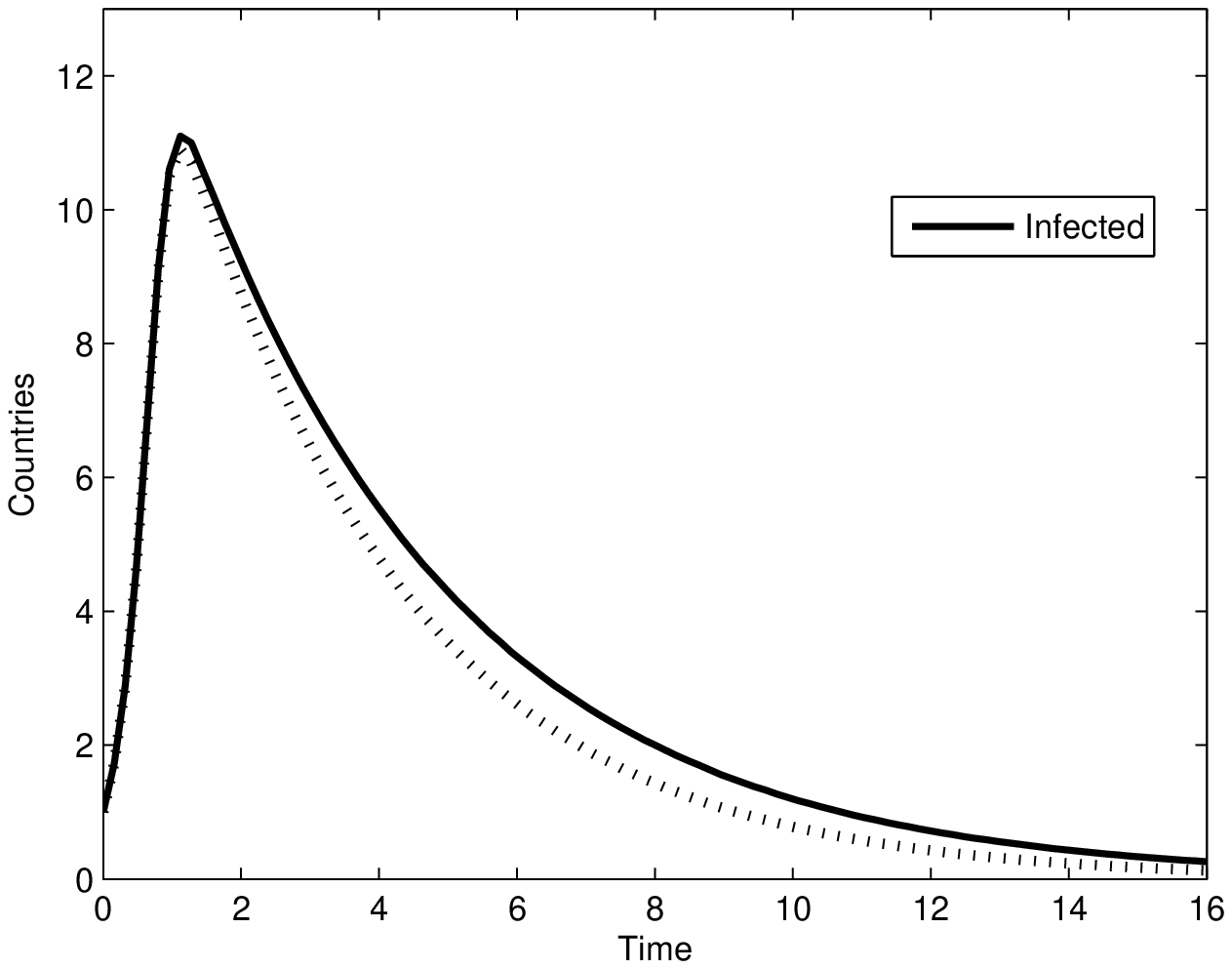}\\
\includegraphics[scale=.42]{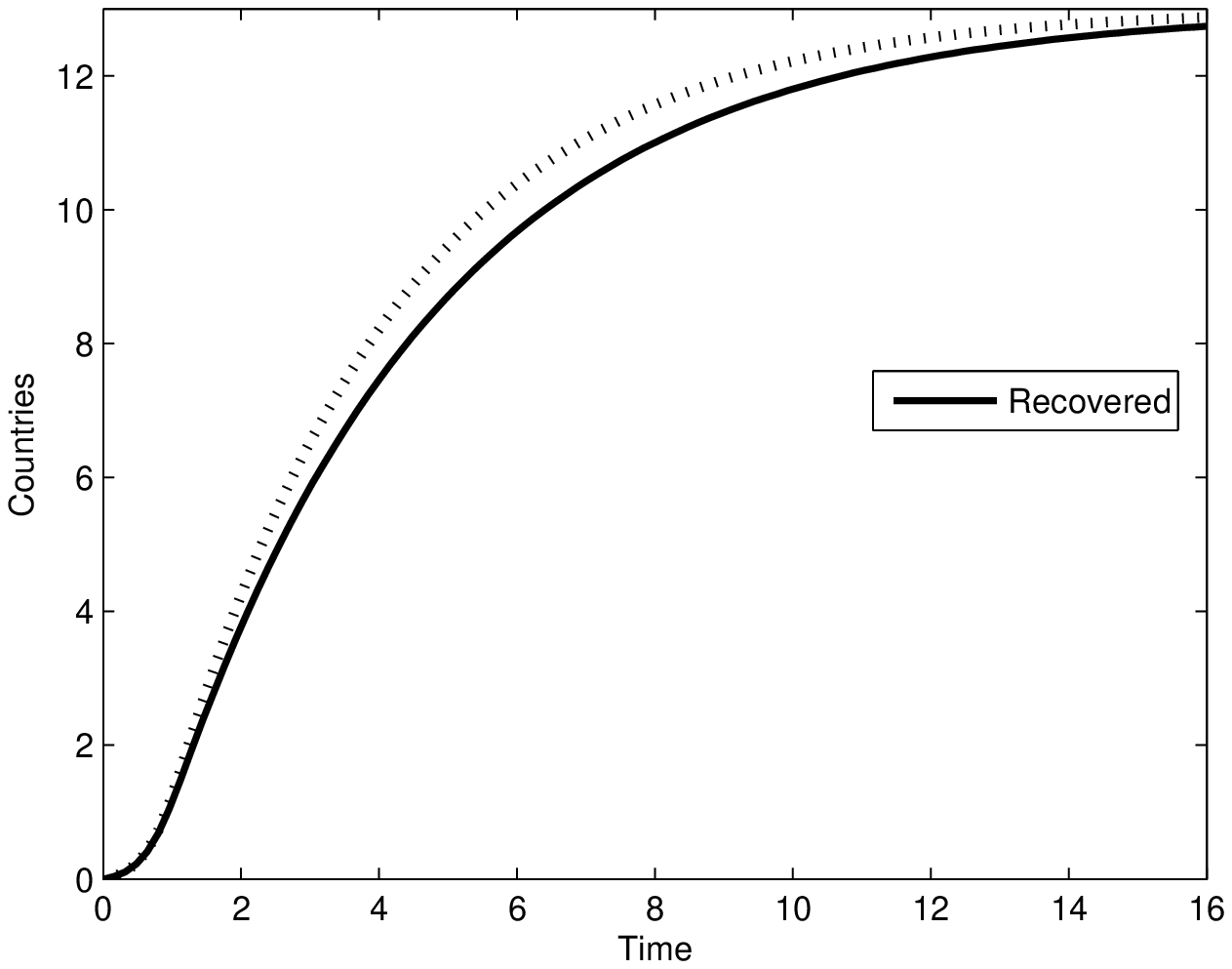}
\includegraphics[scale=.42]{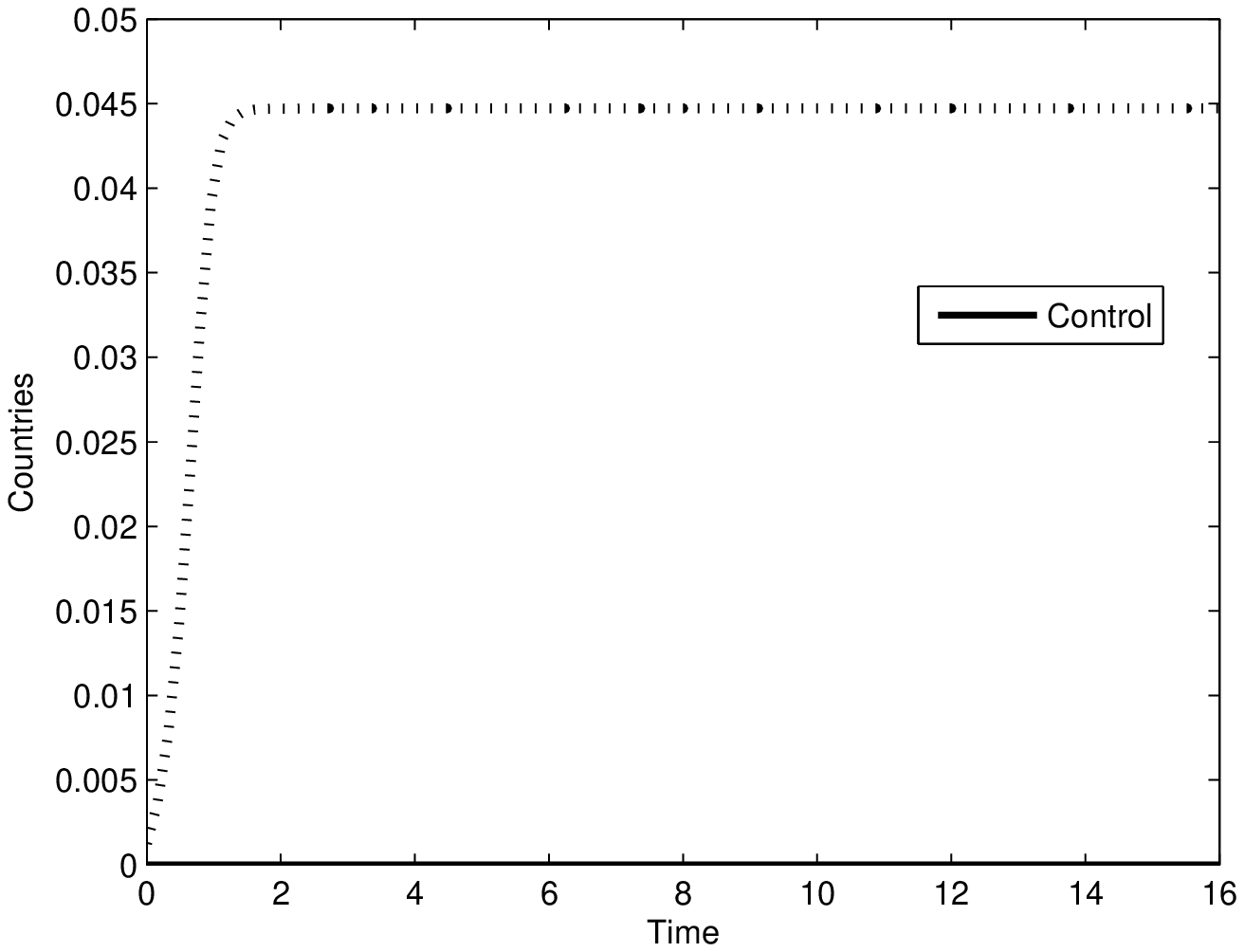}
}
\caption{Contagion risk from USA (Scenario 3): with optimal control 
(dotted line) vs without control (continuous line).}
\label{UsaControl}
\end{figure}

The simulation results are shown in Figures~\ref{PtControl}--\ref{UsaControl}.
They consist of the behavioural curves of our SIR contagion risk model and demonstrate 
the behaviour of the curves with and without optimal control, for different scenarios.

Figure~\ref{PtControl} shows the behaviour of the curves before applying 
the optimal control and after, for the first scenario, when the infection begins in Portugal.
Comparison of the graphics shows that the process of damping of the infection will occur 
almost twice as fast if optimal control is applied, that is, the recovery process 
for countries after financial contamination will occur faster if they receive financial 
support in an optimal way.

The results for the second scenario, that are presented in Figure~\ref{FrControl}, 
confirm the similarity of the conclusions that were formulated for the first scenario.
The infection process will slow down and, simultaneously, the recovery accelerates.
Moreover, the application of optimal control helps more countries to remain 
as susceptible.

Figure~\ref{UsaControl} shows the behaviour of the curves for the third scenario.
We see that the force of infection is quite high and that the application 
of optimal control slightly helps to accelerate the recovery process, 
which leads to a damping of infection.



\section{Conclusion}
\label{sec:06}

We investigated the process of financial contamination from the point of view
of infection spreading among countries that form an interconnected closed-form system.
It was shown how strongly the behaviour of infection depends on the values 
of the transmission and recovery rates, respectively parameters $\beta$ and $\gamma$. 
Interestingly enough, the results obtained by the continuous-time 
epidemiological model \eqref{eqSIR}, which are usually valid for a large $N$ 
(large population), are in agreement with those obtained by the discrete-time (network)
epidemiological model of Section~\ref{sec:04}, which are usually used 
for a small population $N$. The application of optimal control has justified itself, 
showing how significant and successful the results can be on the way to recovery,
thus demonstrating the need for a thorough control of the quantity guaranteed loans
and the financial stability of countries in order to avoid the negative effects of financial risks.
The obtained infection-recovery process corroborates all sorts of macroeconomic 
bibliography premises and therefore validates existing literature 
within economical-cycle expectations. 


\section*{Acknowledgement}

This research was supported by the Portuguese Foundation 
for Science and Technology (FCT) within project 
UID/MAT/04106/2019 (CIDMA). Olena Kostylenko was also supported 
by the Ph.D. fellowship PD/BD/114188/2016. The authors would like 
to thank two reviewers for their critical remarks and
precious suggestions, which helped them to improve the quality 
and clarity of the manuscript.



\end{document}